\documentclass[prd,aps,showpacs,preprintnumbers]{revtex4}

\usepackage{amsmath}
\usepackage{epsfig}
\usepackage{amssymb}
\input{epsf}
\def \be  {\begin{equation}}
\def \ee  {\end{equation}}
\def \ba  {\begin{eqnarray}}
\def \ea  {\end{eqnarray}}
\def \baa {\begin{eqnarray*}}
\def \eaa {\end{eqnarray*}}
\def \bb  {\begin{thebibliography}}  
\def \eb  {\end{thebibliography}}

\def \lab #1 {\label{#1}}

\def \fracs #1#2 {\mbox{\small $\frac{#1}{#2}$}}

\def \bin #1#2 {{\left({#1}\atop{#2}\right)}}
\def \as {\relax\ifmmode\alpha_s\else{$\alpha_s${ }}\fi}

\def \al #1 {\frac {\as({#1})}{\pi} }
\def \ds #1 {\ooalign{$\hfil/\hfil$\crcr$#1$}}

\newcommand \bea{\begin{eqnarray}}
\newcommand \eea{\end{eqnarray}}

\newcommand \ep {\epsilon}
\newcommand \vep {\varepsilon}

\def\hepph  #1 {{\tt hep-ph/#1}}

\begin{document}

\preprint{YITP-SB-07-32}
\preprint{ANL-HEP--PR-07-71}
 
\title{Color Transfer Enhancement for Heavy Quarkonium Production}

\author{Gouranga C.\ Nayak$^{a,b}$, Jian-Wei Qiu$^{c,d}$ 
        and George Sterman$^a$}

\affiliation{
${}^a$C.N.\ Yang Institute for Theoretical Physics,
Stony Brook University, SUNY, 
Stony Brook, New York 11794-3840, U.S.A.\\
${}^b$Department of Physics, University of Illinois, Chicago, IL 60607, USA\\
${}^c$Department of Physics and Astronomy,
      Iowa State University,
      Ames, IA 50011, U.S.A.\\
${}^c$High Energy Physics Division,
      Argonne National Laboratory,
      Argonne, IL 60439, U.S.A.\\
}

\begin{abstract}
We study the transfer of color between a heavy quark pair
and an unpaired heavy quark or antiquark moving at a nonrelativistic
velocity with respect to the pair.  We find that the open heavy quark
or antiquark can catalyze the transformation of the pair 
from octet representation at short distances to singlet
at long distances.   This process is infrared sensitive in general,
   and we exhibit
double poles in dimensional regularization
at next-to-next-to-leading order in the
transition probability.   Because of their
dependence on kinematic variables, these poles cannot be
matched to the non-perturbative matrix elements of effective field
theories based on a single heavy quark pair.
\end{abstract}

\date{\today}
 

\maketitle

\section{Introduction}

Heavy quarkonium production is of special interest 
in quantum chromodynamics
(QCD), because the formation of the heavy quark pair leading up to
its hadronization can be perturbative.
The subsequent evolution of the  pair
has been extensively and fruitfully treated in the language of
effective theories.
Prominent among these are non-relativistic QCD (NRQCD) 
\cite{bodwin94,Brambilla:2004wf,Brambilla:2004jw,Lansberg:2006dh}
and its extensions \cite{Brambilla:1999xf,Luke:1999kz,Hoang:2002ae}.
In NRQCD, the non-perturbative dynamics
is organized through matrix elements of operators
 that are characterized by
an expansion in 
 $v$, the relative velocity of the pair.
 
The NRQCD expansion in $v$ is well-understood for heavy-quarkonium
decays \cite{bodwin94}, although it still lacks a fully compelling 
proof for production processes.
The NRQCD expansion, however, should not be expected to
apply to reactions in which three
heavy particles, two quarks and an antiquark for
example, are produced close together in phase 
space,  because in such cases there is no unique choice
of relative velocity $v$.   In this paper, we will 
extend the reasoning of Ref.\ \cite{Nayak:2007mb}, and argue
that in this limited region of phase space, the formation
of color-singlet quark pairs may be enhanced by a
process that we will refer to below as color transfer.

The application of NRQCD to
production processes has had many successes
\cite{Brambilla:2004wf,Brambilla:2004jw,Lansberg:2006dh,cdf}. 
There are, however, problematic cases.
One involves the polarization of high-$p_T$ charmonia as observed
at the Tevatron \cite{cdfpolar,thypolar}.   
In connection with these data, the importance of
associated production within the
formalism of leading order (LO) NRQCD has
recently been shown  in Ref.\ \cite{Artoisenet:2007xi}.
Another unexpected observation was
the large production cross section 
for $J/\psi$ in association with $c\bar c$ pairs, 
as seen by the BELLE and BABAR collaborations 
\cite{associated}.   It has been shown that
much of this excess may be accounted for
within NRQCD by computing next-to-leading order (NLO) corrections  
\cite{Zhang:2006ay}.

The discussion below will explore the possibility that
a complete picture for associated production will require
us to extend NRQCD itself.
We will argue that  associated production
with $J/\psi$ and related quarkonium states
may be sensitive to
the kinematic region mentioned above, where
heavy quarkonium is produced close in phase
space to open heavy flavor.  The additional quark, 
itself a source of color, could influence
the hadronization  of  these states.  
This mechanism also sheds light
on NRQCD factorization \cite{bodwin03}
for processes without associated heavy quark
pairs, and we will make contact with
our previous work on that important question 
\cite{Nayak:2005rt,Nayak:2006fm}.

In the following section, we begin with a discussion
of infrared poles in dimensional regularization for NLO corrections to
associated production in $\rm e^+e^-$ annihilation  
\cite{Zhang:2006ay}.  We will exhibit 
infrared divergences that, although purely imaginary at this order,
do not match with NRQCD operator matrix elements.
These infrared poles are associated with the transfer
of color between two heavy quarks and an antiquark,
and cannot be associated with a quarkonium wave function.
In Sec.\ 3, we analyze these NLO corrections in
the limit that the relative velocity for
one quark-antiquark combination is much smaller than 
that of the other, while both velocities
are non-relativistic. In this non-relativistic, velocity-ordered region,
we will exhibit an infrared divergence with a
characteristic kinematic enhancement.  
We will go on to identify the effective
nonlocal operator that generates the leading contribution,
and relate it to a similar combination encountered 
in the effective theory known as
potential NRQCD \cite{Brambilla:1999xf}.  

Section 4 treats this color-transfer effect at next-to-next-to-leading
order (NNLO) in the velocity-ordered region.  
We find in this case a real contribution to the
cross section with a double infrared pole.   Generalizing this
purely perturbative result, in Sec.\ 5 we motivate an order-of-magnitude
estimate for the color-transfer contribution to the cross section.
We provide numerical estimates at B factory energies, which
suggest that color transfer can be competitive with the
familiar color singlet mechanism for associated production.

We go on in Section 6 to study briefly the high energy behavior
of the color transfer mechanism.   Although we use the example of
$\rm e^+e^-$ annihilation, many of our observations
apply as well to fragmentation in
hadronic collisions.  We exhibit why, as already
found in \cite{Artoisenet:2007xi}, associated production
is leading power in transverse momentum for the low-relative velocity 
region that we have in mind.   We also touch on issues of polarization
at high energy.   We conclude with a summary
that suggests directions for future work.

\section{NRQCD Matching and NLO Corrections}

Nonrelativistic QCD organizes production and decay
probabilities in terms of an expansion in the relative
velocity, $v\ll 1$, of the quark-antiquark pair that 
eventually forms the bound state.   
In associated production, final states have two pairs.
We will refer to the pair that eventually forms the bound
state as the ``active" pair.   We will refer to the 
additional heavy particle that is closest in phase
space to the active pair as the ``spectator"
quark (or antiquark).   

The quark and antiquark momenta of the
active pair are  $P_1$ and $P_2$, respectively.
These momenta are related to the total and relative momenta
and their rest-frame relative velocity by
\bea
P_1=\frac{P}{2}+q\, , \quad P_2 = \frac{P}{2}-q\, , 
\quad v^2=\frac{q^2}{E^*{}^2}
= 1-\frac{4m^2}{P^2}\, .
\label{Pqvdefs}
\eea
Here $2E^*$ is the total energy of the active
pair in the pair rest frame, and the
second expression for $v$ gives its relation to
the quark mass and the invariant mass of the pair.

Our discussion in this section will lead to a direct evaluation of 
familiar one-loop diagrams in the eikonal approximation,
which will be followed in a later section by an analysis of 
the infrared behavior of all relevant two-loop corrections.
In essence, we will supplement below the calculations
of Refs.\ \cite{Nayak:2005rt,Nayak:2006fm}, by studying 
infrared poles in corrections to the production of a heavy
quark pair due to the exchange of soft gluons with
an additional {\it massive} final-state particle.   The
extension from massless to massive particles in the final
state has important consequences for
a cross section in which the color of the quark pair
is fixed to be in a singlet configuration in the final state.
When other massive particles are observed,
double infrared poles in dimensional regularization
that are absent in the massless case appear 
in NRQCD coefficient functions.   These poles 
have residues that depend sensitively
on the relative momenta of the pair and the 
heavy particles.   They are sharply peaked at 
small relative momentum, but fall off very rapidly
when the relative momentum exceeds the 
masses.

\subsection{Lowest order and NRQCD}

\begin{figure}[t]
\begin{center}
\begin{minipage}[c]{1.4in}
\epsfig{figure=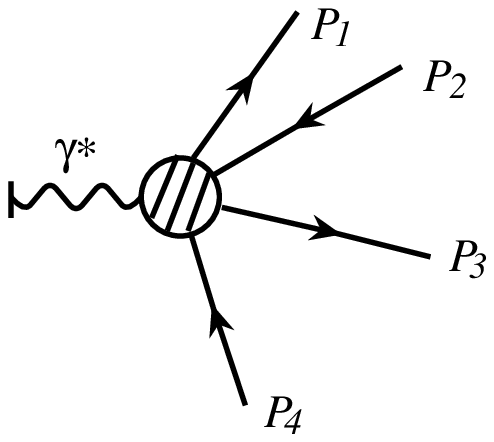,width=1.4in}
\end{minipage}
\hskip 0.2in
=
\hskip 0.2in
\begin{minipage}[c]{1.4in}
\epsfig{figure=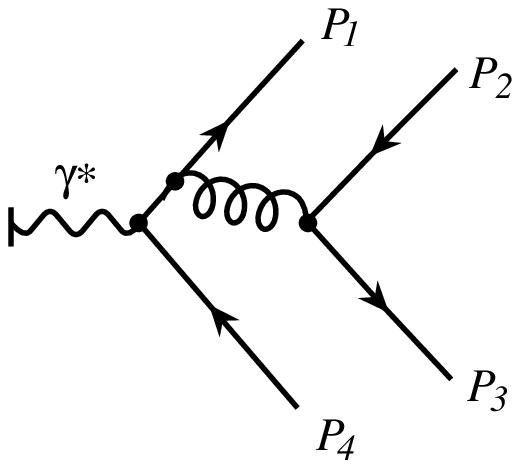,width=1.4in}
\end{minipage}
\hskip 0.2in
+
\hskip 0.2in
\begin{minipage}[c]{1.4in}
\epsfig{figure=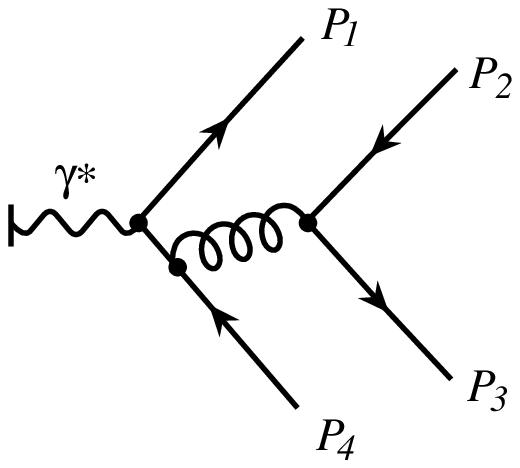,width=1.4in}
\end{minipage}

+
\hskip 0.2in
\begin{minipage}[c]{1.4in}
\epsfig{figure=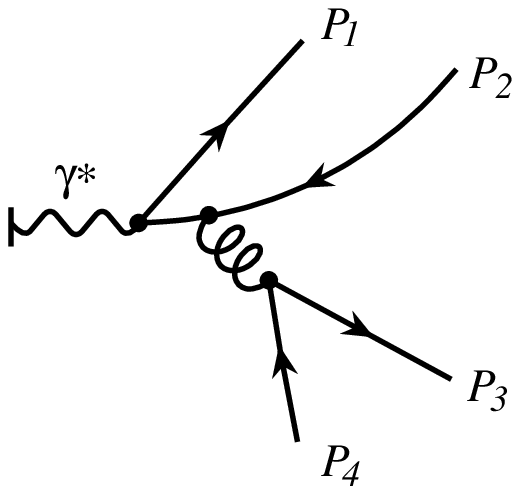,width=1.4in}
\end{minipage}
\hskip 0.2in
+ 
\hskip 0.2in
\begin{minipage}[c]{1.4in}
\epsfig{figure=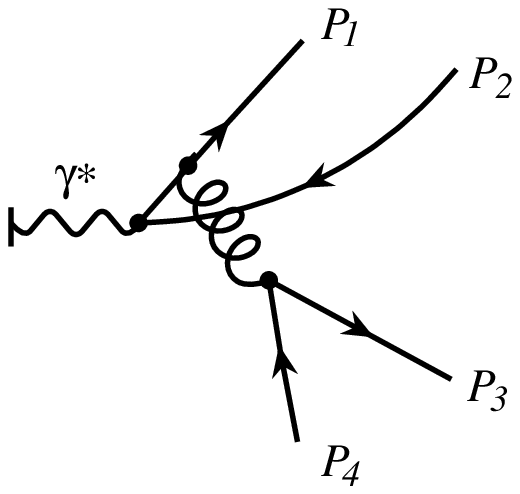,width=1.4in}
\end{minipage}
+
\hskip 0.2in
\begin{minipage}[c]{1.4in}
\epsfig{figure=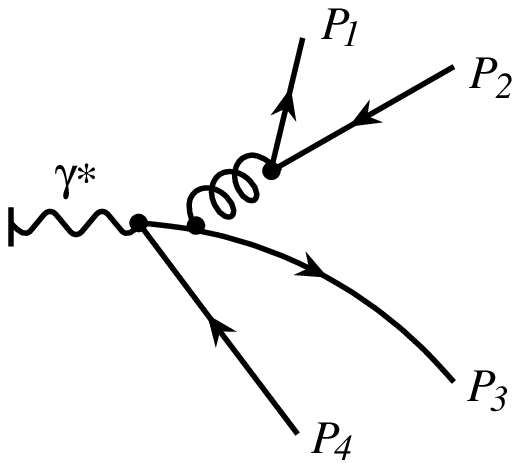,width=1.4in}
\end{minipage}
\hskip 0.2in

+ 
\hskip 0.2in
\begin{minipage}[c]{1.4in}
\epsfig{figure=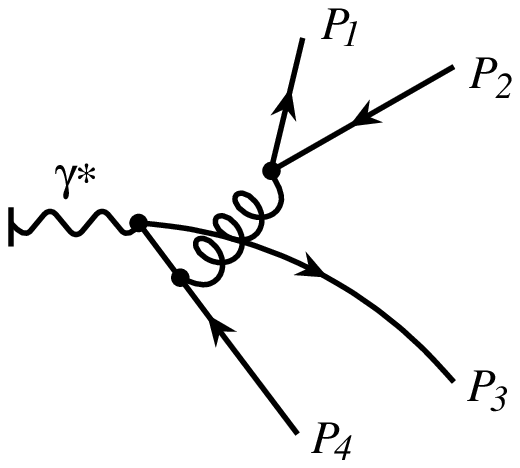,width=1.4in}
\end{minipage}
+
\hskip 0.2in
\begin{minipage}[c]{1.4in}
\epsfig{figure=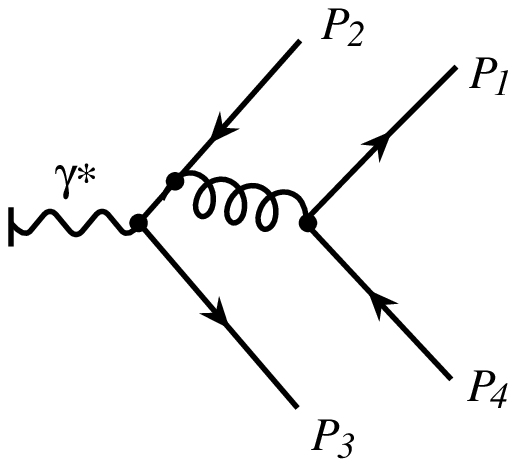,width=1.4in}
\end{minipage}
\hskip 0.2in
+ 
\hskip 0.2in
\begin{minipage}[c]{1.4in}
\epsfig{figure=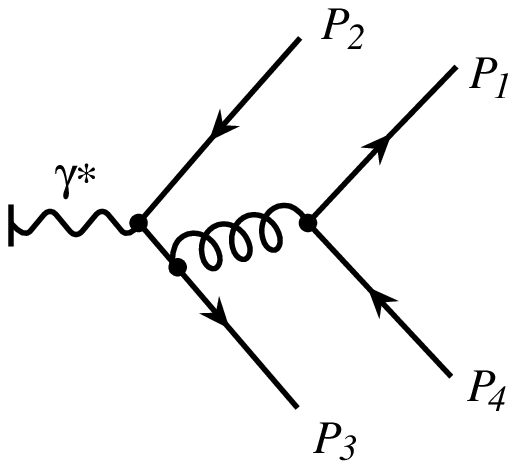,width=1.4in}
\end{minipage}

\caption{The LO diagrams for
a virtual photon decaying into two charm quark pairs, at
fixed quark momentum  $P_1$ and antiquark momentum $P_2$.
The pair $P_1$ and $P_2$ will be associated with a bound state.
}
\label{lofig}
\end{center}
\end{figure}

Figure \ref{lofig} shows the lowest-order diagrams that 
contribute to associated production in 
$e^+e^-\rightarrow Q\bar{Q}+Q\bar{Q}$ annihilation
through a virtual photon.  To form a heavy
quarkonium state, it is assumed that the members of the active
pair are close together in phase space,
with relative velocity $v\ll 1$ in Eq.\ (\ref{Pqvdefs}).

To derive the NRQCD expression for the production
of heavy quarkonium $H$ 
corresponding to
the diagrams of Fig.\ \ref{lofig}, 
we first project these amplitudes onto  a complete set of spin,
angular momentum and color states, collectively labeled $n$, for
the pair.  We then square these amplitudes, normalize appropriately, 
and multiply each of the resulting expressions
by the corresponding NRQCD matrix elements, which
summarize nonperturbative dynamics.  Summing over $n$, we
have, schematically,
\begin{equation}
d\sigma_{e^+e^-\to  H+X}(p_H) = 
\sum_n\; d\hat\sigma^{\rm  (LO)}_{e^+e^-\to  Q\bar{Q}[n]+X}(p_H)\,
\langle 0|\, {\mathcal  O}^H_n\, |0\rangle\, ,
\label{nrfact}
\end{equation}
with $p_H$ the quarkonium momentum.
Color singlet NRQCD operators are of the general form
  \cite{bodwin94} 
\ba
{\mathcal O}^H_{n(1)}(0)
&=&
\sum_N\ \chi^\dagger(0){\mathcal \kappa}_{n(1)}\psi(0)\, 
\left | N,H\right\rangle\,
\left\langle N,H \right|\, \psi^\dagger(0){\mathcal \kappa}'_{n(1)}\chi(0)
\nonumber\\
&=&
\chi^\dagger(0){\mathcal \kappa}_{n(1)}\psi(0)\, 
\left(a^\dagger_Ha_H\right)\,
\psi^\dagger(0){\mathcal \kappa}'_{n(1)}\chi(0)
\, ,
\label{Ondef1}
\ea
with $\kappa_n$ and $\kappa'_n$
projections for color (octet or singlet) and spin.
For this factorization to be useful, it is necessary that 
the ``coefficient function" $d\hat\sigma_{e^+e^-\to  Q\bar{Q}[n]+X}(p_H)$ 
be infrared safe at higher orders.
This factorization remains a conjecture for production processes
beyond NNLO. 

In Ref.\ \cite{Nayak:2005rt} we showed that, beyond NLO, factorization
requires a reformulation of octet production matrix elements into a 
 ``gauge-completed" form,
\ba
{\mathcal O}^H_{n(8)}(0)
&=&
\chi^\dagger(0){\mathcal \kappa}_{n(8),c}\psi(0)\, 
\Phi_l^{(A)}{}^\dagger(\infty,0){}_{cb}\, 
\left(a^\dagger_Ha_H\right)\,
\Phi_l^{(A)} (\infty,0)_{ba}\, \psi^\dagger(0) 
{\mathcal \kappa}'_{n(8),a}\chi(0)\, ,
\nonumber\\
\Phi_l^{(A)}(\lambda_1,0) &\equiv& {\cal P}\ \exp 
\left[\, -ig\int_0^{\lambda_1} 
d\lambda\; l^\mu A_{\mu,a}(\lambda l)\, T_a^{(A)}\, \right]\, .
\label{complete}
\ea
Here, as shown, $\Phi_l^{(A)}(\lambda,0)_{ab}$ 
is a gauge link in the adjoint representation ($a,b=1\dots 8$)
with a light-like four-velocity $l$,
and with ${\cal P}$ path ordering for the expansion in $g$.
We showed, in particular, that gauge-completed matrix elements
are independent of the direction of the gauge link, so
long as $l$ remains light-like.  We will see
 in the calculations that follow, however, that 
 independence of the direction of $l$ does not
 generalize to a massive color source 
 on the right-hand side of Eq.\ (\ref{complete}).   This observation
 will have important consequences for associated production.
 To see how this comes about, we turn now to 
 soft gluon loop corrections to the lowest order diagrams
 of Fig.\ \ref{lofig}. 

\subsection{NLO soft gluon corrections and matching}

It would be a major undertaking to compute the full NLO corrections to
the singlet NRQCD cross section for two pairs
at arbitrary momenta.  Nevertheless, it is relatively
straightforward to check the self-consistency of the NRQCD 
factorization at NLO in this process.  This can be done by checking that
 infrared poles in dimensional regularization either cancel,
 or can be absorbed into vacuum expectation values
 $\langle 0|{\cal O}_n|0\rangle$, thus matching
 full QCD to NRQCD.  Matching is essentially
 equivalent to NRQCD factorization.
 
  More specifically, NRQCD
factorization and  matching for production
cross sections require the cancellation
of all infrared gluons that are not ``topologically factorized" 
into factors equivalent to the perturbative expansions of matrix elements in
the effective theory \cite{bodwin94,Nayak:2005rt}.   
The allowed, topologically factorized, 
soft gluons are those that can be absorbed 
either into
the interaction of the active pair with ``the vacuum",
or that do not couple to either of the 
heavy quarks that form the quarkonium.  In the former
case, the soft gluons generically will have the interpretation
of part of a nonperturbative matrix element or wave function.  
In the latter case, they will cancel in the inclusive 
cross section for fixed active pair color representation.

\begin{figure}[h]
\begin{center}
\begin{minipage}[c]{1.4in}
\epsfig{figure=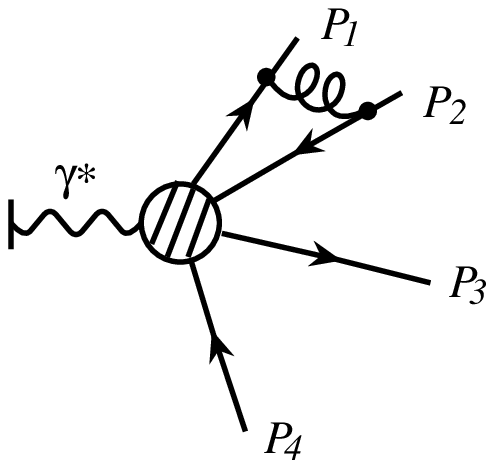,width=1.4in}
\end{minipage}
\hskip 0.2in
+
\hskip 0.2in
\begin{minipage}[c]{1.4in}
\epsfig{figure=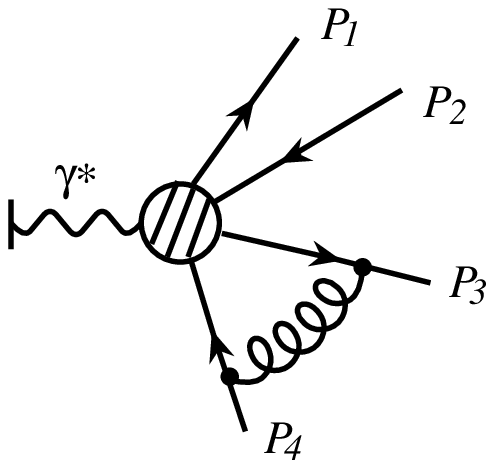,width=1.4in}
\end{minipage}

+
\hskip 0.2in
\begin{minipage}[c]{1.4in}
\epsfig{figure=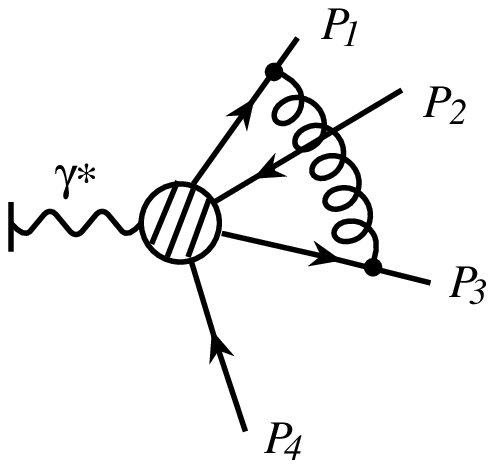,width=1.4in}
\end{minipage}
\hskip 0.2in
+ 
\hskip 0.2in
\begin{minipage}[c]{1.4in}
\epsfig{figure=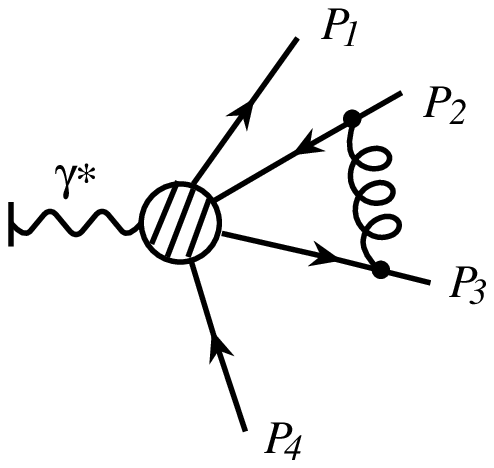,width=1.4in}
\end{minipage}

\caption{Diagrams for the one-loop virtual infrared corrections, 
where the blob represents a complete set of lowest order
Feynman diagrams, as shown in Fig.~\protect\ref{lofig}.
}
\label{nlofig}
\end{center}
\end{figure}

Figure \ref{nlofig} illustrates these considerations.
With  $P_1$ and $P_2$
the momenta of the active pair, 
the diagram on the upper left of the figure shows
 a virtual soft gluon that
is emitted from the active quark and absorbed by
the active antiquark.    Such a correction
can be absorbed into a wave function.
Similarly, contributions that describe the
interference between gluon emission by the active
quark in the amplitude with
emission by the active antiquark in the complex
conjugate amplitude are also topologically factorized,
and hence consistent with NRQCD matching.

The right of the upper line of Fig.\ \ref{nlofig} 
shows a gluon that connects to
neither of the active quarks.  Such a contribution cancels
in the sum over final states at fixed color representation
of the active quarks.   It is thus also consistent with NRQCD matching.

In the remaining two diagrams of Fig.\ \ref{nlofig},
a member of the pair interacts with a spectator quark,
of momentum $P_3$ in this case.   
These two diagrams are therefore not topologically factorized.
Nevertheless, in a production cross section the
real infrared poles of these diagrams still cancel
against diagrams with soft gluons in the final state,
for a fixed color projection on the active quark pair \cite{bodwin94}.
Of course any imaginary terms cancel 
upon combination with complex conjugate diagrams.
At NLO, therefore, the non-topologically factored 
diagrams do not affect NRQCD
calculations \cite{Zhang:2006ay}.

In Ref.\ \cite{Nayak:2006fm}, however,
we found a somewhat stronger result for the
imaginary poles of the non-factored diagrams of Fig.\ \ref{nlofig}.  
When the spectator momentum is light-like ({\it i.e.}, $P_3^2=0$),
the ``Coulomb phase" associated with the
exchange of a soft gluon between a massive quark
and a massless quark is independent of their relative
directions.   As a result, the imaginary infrared poles of 
the two non-factored diagrams 
in Fig.\ \ref{nlofig}
differ only by a relative minus sign between
the quark and antiquark in the active pair.  
Once we project on a singlet final color state for the pair,
they then cancel identically,
to {\it all} orders in the relative velocity, $v$, Eq.\ (\ref{Pqvdefs}).
This is an important ingredient in the NNLO factorization
of gauge-completed octet NRQCD matrix elements,
Eq.\ (\ref{complete}).

We will reproduce these diagrammatic
results below, but show that they no longer hold
for a massive spectator (anti)quark.  
For a massive spectator, the infrared pole 
is a simple but nontrivial function of the pair relative
velocity $v$, as well as the pair-spectator relative
velocity.

At the level of infrared poles in dimensional regularization,
the one-loop virtual corrections
for the diagrams shown in Fig.~\ref{nlofig} can be 
factorized into LO short-distance color matrices, $M^{\rm (LO)}$, times
long-distance color matrices, which describe the
exchange of soft gluons.  The latter
can be calculated in the eikonal approximation,
in which the $j$th quark or antiquark propagator 
and vertex are replaced by 
the spin-indendent combination
\ba
\pm g \frac{P_j^\mu}{P_j\cdot k}\left(T^{(f)}_a\right)_{i_j;i'_j}\, ,
\label{eikonal}
\ea
with the plus sign for a quark and the minus for an antiquark,
and where $f=q$ or $\bar{q}$ denotes the quark or
antiquark representation for the generators $T^{(f)}_a$.  
Equivalently,
we may take all color matrices in the defining (quark) representation,
and stipulate that matrix multiplication follows the arrows that
represent the flow of fermion number.

\begin{figure}[h]
\begin{center}
\epsfig{figure=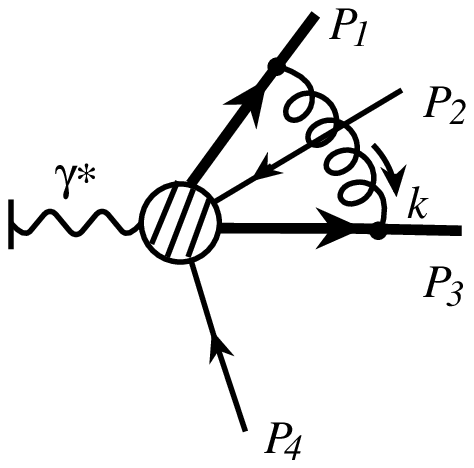,width=1.4in}
\hskip 0.6in
\epsfig{figure=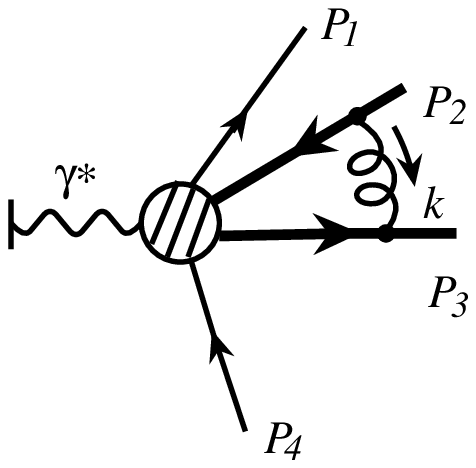,width=1.4in}
\caption{Diagrams with eikonal interactions between the heavy quark
pair and an associated heavy quark of momentum $P_3$.  
The heavy lines indicate fermions in the eikonal approximation.
}
\label{dashedgluon}
\end{center}
\end{figure}

In these terms we write the factorized amplitude as
\ba
A^{\rm (NLO,IR)}_{i_3 i_4}
=
\sum_{i'_1\dots i'_4}\ \delta_{i_1i_2}
{\cal A}^{(1)}_{i_1 \dots i_4;i'_1 \dots i'_4}(P_1,P_2;P_3,P_4)\;
M^{\rm (LO)}_{i'_1\dots i'_4}(P_1,P_2;P_3,P_4)\, ,
\label{factorme}
\ea
where $M^{\rm (LO)}$ is the leading-order amplitude 
projected on the appropriate spin state, at fixed values of 
color indices.  
${\cal A}^{(1)}$ is the eikonal factor, 
describing the coupling of a soft gluon to the
outgoing quarks.  The diagrams of Fig.~\ref{dashedgluon}
are typical contributions to ${\cal A}^{(1)}$, and in this case
correspond to the non-factored diagrams in Fig.\ \ref{nlofig},
with the heavy lines representing the eikonal approximation
(\ref{eikonal}) 
for the couplings of soft gluons of momentum $k$.  

In Eq.\ (\ref{factorme}),  color indices $i_1$ and $i_2$ correspond to the 
active pair, of momenta $P_1$ and $P_2$, which are
associated with the heavy quarkonium.
As shown, we set $i_1=i_2$ and sum to enforce
a color singlet final state.  
The remaining color indices $i_3$ and $i_4$ 
correspond to the open heavy quark pair, of momenta $P_3$ and $P_4$, 
as shown in Fig.~\ref{lofig}.  We now turn to the evaluation of the
non-factored contributions.

\subsection{Infrared poles in the NLO amplitude}

The contributions of the two diagrams in Fig.~\ref{dashedgluon} 
to the eikonal factor in Eq.\ (\ref{factorme}) are given 
in $D=4-2\varepsilon$ dimensions by
\begin{eqnarray}
\left[{\cal A}_{13}\right]_{i_1\dots i_4;i'_1\dots i'_4}
&=& -i\, g^2\,\mu^{2\vep}\;
\left(T^{(q)}_a\right)_{i_1i'_1}\left(T^{(q)}_a\right)_{i_3i'_3}\ 
\delta_{i_2i'_2}\delta_{i_4i'_4}
\nonumber \\
&\ & \hspace{10mm} \times
\int \frac{d^D k}{(2\pi)^D}\
\frac{P_1\cdot P_3}
     {[P_1\cdot k + i\epsilon]\,
      [-P_3\cdot k + i\epsilon]\,
      [k^2 + i\epsilon]} \, , \nonumber\\
\left[{\cal A}_{23}\right]_{i_1\dots i_4;i'_1\dots i'_4}
&=& -i\, g^2\,\mu^{2\vep}\;
\left(-T_a^{(q)\, *}\right)_{i_2i'_2}\, 
\left(T^{(q)}_a\right)_{i_3i'_3}\ 
\delta_{i_1i'_1}\delta_{i_4i'_4}
\nonumber \\
&\ & \hspace{10mm} \times
\int \frac{d^D k}{(2\pi)^D}\
\frac{P_2\cdot P_3}
     {[P_2\cdot k + i\epsilon]\,
      [-P_3\cdot k + i\epsilon]\,
      [k^2 + i\epsilon]} \, ,
\label{sf_nlo-b}
\end{eqnarray}
with ${\cal A}={\cal A}_{13} + {\cal A}_{23}+\dots$.  
For the analysis of this section, we will assume
equal masses $m=\sqrt{P_1^2}=\sqrt{P_2^2}$ for the 
active quark pair, but allow the additional
heavy flavor to have a possibly different but nonzero mass, 
$m_3=\sqrt{P_3^2}$.

In accordance with our previous discussion, we will be interested
in the relative motion of the three heavy particles.
For the equal-mass case, a convenient choice for relative velocity
between $P_i$ and $P_j$ is $\beta_{ij}=\sqrt{1-4m^2/s_{ij}}$, 
where we adopt the notation
\bea
s_{ij} = (P_i+P_j)^2\, .
\eea 
As in Eq.\ (\ref{Pqvdefs}), we define $v=\beta_{12}$ for the relative
velocity between the active pair. 
Consider now momenta $p$ and $p'$ with arbitrary masses, $m$ and $m'$.
Their relative velocity in any frame where their
spatial components are collinear is given by
\bea
\beta(p,p') =  \left | \bf{v}-\bf{v}' \right |_{\mathrm col}
&=&
\frac{1}{EE'}\ \sqrt{(p\cdot p')^2-(mm')^2}\, ,
\label{betadef}
\eea
where $E=p_0$, and similarly for $E'$ in the same frame.
For equal masses and in the center of mass frame, $\beta(P_i,P_j)$
reduces to twice the familar relative velocity, $\sqrt{1-4m^2/s_{ij}}$,
of Eq.\ (\ref{Pqvdefs}).

Another measure of the distance in phase
space between two vectors that arises
in the soft gluon corrections at hand is
\bea
\bar\beta(p,p') = \frac{EE'}{p\cdot p'}\; \beta(p,p')
=
\sqrt{1-\frac{m^2m'\, {}^2}{(p\cdot p')^2}}\, ..
\label{barbetadef}
\eea
We easily verify the limiting behaviors,
\bea
\lim_{(mm')/(p\cdot p') \to 1} \bar\beta(p,p') = \beta(p,p')\, ,
\nonumber\\
\lim_{(mm')/(p\cdot p') \to 0} \bar\beta(p,p') = 
\frac{1}{2}\; \beta(p,p')\, .
\eea
That is, for small relative velocities, $\bar\beta$ is nearly
equal to $\beta$, although it increases somewhat more slowly
with center-of-mass momenta, reaching a limit 
of $\beta\sim 1$ for fully relativistic motion.
We will be interested in regions of phase space
for which relative velocities are nonrelativistic.  In
such regions, $\bar\beta, \beta \ll 1$.   For convenience, we will  
refer to $\bar\beta$ as a velocity.

So that we may unambiguously distinguish
the spectator quark, we restrict ourselves to momenta for which 
$v < \bar\beta\ll 1$. 
In this region of phase space, conventional
NRQCD factorization need not apply directly,
because of the presence of an additional small parameter, $\bar\beta$.
Indeed, we will find  an additional infrared-sensitive enhancement in
the octet-to-singlet conversion for the active
quark pair, due to the presence of a nearby spectator.
If this mechanism is
observable, it should manifest itself in a peak in 
associated open heavy flavor distributions in phase space near quarkonia.

The evaluation of each integral in Eq.\ (\ref{sf_nlo-b}) is 
reasonably straightforward, and we readily isolate their
infrared poles.  In fact, in dimensional regularization
these integrals vanish through the cancellation of IR
and UV poles.   Nevertheless, the single infrared pole
may be found (for example)  by adding quadratic terms $+k^2/2$
to the eikonal denominators.
The result is
\bea
&& {\rm Pole}^{\rm (IR)}\ \left[\, -ig^2\, \int \frac{d^D k}{(2\pi)^D}\
\frac{P_i\cdot P_j}
     {[P_i\cdot k + i\epsilon]\,
      [-P_j\cdot k + i\epsilon]\,
      [k^2 + i\epsilon]}\, \right] 
      \nonumber\\
      && \hspace{30mm}
= -\, \frac{1}{2\varepsilon}\, \frac{\alpha_s}{2\pi}   
\frac{1}{\bar \beta(P_i,P_j)} 
\left( \ln \left [ \frac{ 1+ \bar\beta(P_i,P_j)}
                        {1-\bar\beta(P_i,P_j)} 
\right ] - 2i \pi   \right)\, ,
\label{nlointegral}
\eea
where $\bar\beta$ has been defined in (\ref{barbetadef}).
In the region that we have identified above, 
$ \bar\beta(P_i,P_j) \ll 1$, we
observe that for the infrared pole, singular behavior
in $\bar\beta$ appears in the imaginary, but not in the real part, 
\bea
&& {\rm Pole}^{\rm (IR)}\ 
\left[\, -ig^2\, \int \frac{d^D k}{(2\pi)^D}\
\frac{P_i\cdot P_j}
     {[P_i\cdot k + i\epsilon]\,
      [-P_j\cdot k + i\epsilon]\,
      [k^2 + i\epsilon]}\, \right] 
      \nonumber\\
      && \hspace{30mm}
=  -\, \frac{1}{\varepsilon}\, \frac{\alpha_s}{2\pi}   
\frac{1}{\bar \beta(P_i,P_j)} 
\left(  \bar \beta(P_i,P_j) -i \pi   \right)\ 
+ \ {\cal O}(\bar \beta(P_i,P_j))\, .
\eea
Our discussion below will concentrate on the effect of this
pole on production cross sections, which, as we shall show, 
is seen first at NNLO.

At this point we  note that the real part of the 
full vertex correction has a (famous) $1/\bar\beta$ singularity.  This
power singularity near threshold,
however, is finite in four dimensions, and is associated with loop
momenta at the scale of $mv^2$, the so-called ``ultrasoft" momenta.
Dynamics at this scale are regulated
by bound state effects between the active quarks.  
In this paper, we shall
not attempt an analysis of exchanges at the ultrasoft scale between the
active pair and spectatators.

Consider now
the  imaginary contributions of Eq.\ (\ref{sf_nlo-b}) to ${\cal A}$,
traced over the colors of the active pair to enforce
a singlet configuration in the final state,
\ba
\sum_{i_1i_2}\, \delta_{i_1i_2}\ {\rm Im}
\left[{\cal A}_{13}+{\cal A}_{23}\right]_{i_1\dots i_4;i'_1\dots i'_4}
&=& \frac{1}{\vep}\, \left(\frac{\alpha_s}{2}\right) \
\, \left(T_a\right)_{i'_2i'_1}\left(T_a\right)_{i_3i'_3}\ 
\delta_{i_4i'_4}
\nonumber\\
&\ & \hspace{-15mm} \times
\left[
\frac{1}{\sqrt{1-P_1^2P_3^2/(P_1\cdot P_3)^2}}
-
\frac{1}{\sqrt{1-P_2^2\,P_3^2/(P_2\cdot P_3)^2}}
\right]\, .
\label{allorder}
\ea
This pole in the imaginary part vanishes
identically for $P_1\cdot P_3=P_2\cdot P_3$
or for $P_3^2=0$.  In the former case, the relative
velocity of the spectator quark and the active quark
equals the relative velocity of the spectator and
the active antiquark.   The interpretation of the
cancellation is then that soft gluons
emitted by the spectator cannot resolve the
charges of a perfectly co-moving pair
in a singlet color state.
Analogously, when the spectator is 
lightlike, the relative velocity to both active lines is unity,
and the two terms cancel.  Evidently, in
this case a light-like spectator cannot
resolve a color singlet even when the
quark and antiquark are not co-moving.

From Eq.\ (\ref{allorder}), the sum of the exchanges 
between the active pair and the spectator depends directly
on the relative velocities of the spectator with the
active quark and antiquark, but only indirectly
on the relative velocity $v$ of the active pair itself.
In general, for $\beta_{13}\sim \beta_{23}$,
we must describe the kinematics of two quarks vying 
for the favor of a single
antiquark to form a heavy bound state, with no obvious favorite.
This three-body problem is potentially difficult  to analyze.
The soft gluon dynamics of two quarks and an antiquark
might be generated from local operators like
$\psi^\dagger_{P_1}\chi_{P_2}\psi^\dagger_{P_3}$, with
time evolution generated by heavy quark effective
theory for each of the quarks, but only if 
we neglect the recoil of the quarks.
Nonrelativistic QCD organizes such corrections
into nonperturbative matrix elements, but only
to the extent that the extra quark may be neglected.

We will not attempt a full solution to the three-particle
problem, but will concentrate in the following on
a limit more closely related to NRQCD.     We have already noted that 
the sum of NLO phases in Eq.\ (\ref{allorder}) vanishes
in the limit $v=\beta_{12}\to 0$, at fixed, finite values
of $\bar\beta_{13}=\bar\beta_{23}$.   
We now  turn to this region of nonrelativistic,  ordered velocities,
which we will refer to as the ``velocity-ordered region".

\section{NLO Color Transfer for Ordered Velocities}

We begin this section by expanding the 
residue of the infrared pole on the right-hand side of Eq.\ (\ref{allorder}) 
to lowest order in the active pair relative velocity,
$v$, in the velocity-ordered region.   
The residue is strongly peaked
toward low relative velocity for the spectator pair.
We will then go on to rederive this result from
an effective non-local vertex involving the gluon field 
strength coupled to the dipole moment of the 
active pair \cite{Nayak:2005rt}.   This reformulation  will enable us to
extend our analysis to NNLO in the following section,
and also to exhibit the relation of the analysis here
to our previous study of  NRQCD factorization \cite{Nayak:2006fm}
and to potential NRQCD \cite{Brambilla:1999xf}.

\subsection{Expansion in $v$}

To estimate the kinematic behavior of
the poles in Eq.\ (\ref{allorder}) in the velocity-ordered region, 
we expand at low relative velocity, $v$ for the active pair.
In Eq.\ (\ref{allorder}), the only kinematic variables are
$P_1\cdot P_3$ and $P_2\cdot P_3$, and we expand both 
around their values at $v=0$, $(P/2)\cdot P_3$, with $P^2=4m^2$
We will allow $P_3$ to have an arbitrary nonzero mass, $m_3$.
Given that we have two invariants, we will have to introduce
an additional relative velocity, and corresponding relative momentum.

We begin this straightforward kinematic analysis
with Eq.\ (\ref{Pqvdefs}) for the momenta of the active pair,
recalling that the mass-shell conditions $P_1^2=P_2^2=m^2$
require $q\cdot P=0$.
In a frame where the total pair momentum is  at rest, the
relative momentum, $q^\mu$ has no time component.  We can then
write the total energy of the pair as a function of $q$, in
terms of a momentum at rest in that frame, $P_0^\mu=(2m,\vec{0})$.  
In covariant form 
the dependence of $P$ on $q$ is given explicitly by
\bea
P^\mu &=& P_0^\mu\, \sqrt{1 - \frac{q^2}{m^2}}\, ,
\eea
where $q^2= - \vec{q}_{\rm c.m.}^{\ 2}$.
We now parameterize $P_3^\mu$ in a similar way, 
expanding around $P_3=(m_3,\vec{0})$ in
the $P^\mu$ (active pair) rest frame, in terms of
a  vector $q_S^\mu$, which like $q$ has vanishing time
component in that frame,
\bea
P_3^\mu = m_3\; \frac{P_0^\mu}{2m}\ 
\sqrt{1 - \frac{q_S^2}{m_3^2}} + q_S^\mu\, ,
\quad
q_S\cdot P_0 = 0\, .
\eea
Expressing $P_3$ in these terms, and recalling 
that $P_1=P/2+q$, we evaluate 
the relevant invariant $P_1\cdot P_3$ 
as a function of $q$ and $q_S$, finding
\bea
P_1\cdot P_3 &=&   m\, m_3\; 
\sqrt{\left(1 - \frac{q^2}{m^2} \right)
      \left(1 - \frac{q_S^2}{m_3^2}\right) }
 + q\cdot q_S\, .
 \label{P13}
 \eea
The analogous result for $P_2\cdot P_3$ is found 
by simply changing the sign of
$q\cdot q_S$.  We are now ready
to expand (\ref{allorder}) in $q$ at fixed $q_S$.
 
Starting at $q=0$,
$P_1=P_2=P_0/2$,  both invariants  $P_i\cdot P_3$, $i=1,2$ are given by
\bea
 (P_0/2)\cdot P_3 =  m\, \sqrt{m_3^2 - q_S^2} \equiv   
 \frac{m\, m_3}{\sqrt{1-\bar\beta_S^2}}\, ,
\label{P0cdotP3}
\eea
where in the second expression we recall the
definition of the function $\bar\beta$, Eq.\ (\ref{barbetadef}), 
and define
\bea
 \bar\beta_S \equiv \bar\beta(P_0/2,P_3) 
= \sqrt{\frac{-q_S^2}{m_3^2 - q_S^2}}\, .
\label{betaSdef}
\eea
This is a measure of the spectator's velocity relative to
the active pair, neglecting the latter's internal relative veloctiy, $v$.
We note that this particular velocity is independent of $m$, 
so that, for example, $ \bar\beta(P_0/2,P_3) = \bar\beta(P_0,P_3)$,
and  $\bar\beta_S=1$ for  $m_3=0$.
 
Following our previous discussion, we consider the region $q^2 < q_S^2$,
and expand in the corresponding ratios on the right-hand side of 
Eq.\ (\ref{allorder}),
keeping in mind that at $q=q_S=0$ both terms in the difference 
on the right-hand side diverge.    
We note as well that only terms that are odd in $q^\mu$ survive
 in the difference.   We find
\bea
\frac{1}{\sqrt{1-P_1^2P_3^2/(P_1\cdot P_3)^2}}
-
\frac{1}{\sqrt{1-P_2^2\,P_3^2/(P_2\cdot P_3)^2}}
= 
\frac{2}{ \bar\beta^3_S}\ \frac{q_S\cdot q}{m\, m_3}\; 
\left (1 - \bar\beta_S^2 \right )^{3/2}\
\left[\, 1+ {\cal O}\left(\frac{q^2}{m^2} \right)\, \right]\, .
\nonumber\\
\label{betaminusthree}
\eea
This result is exact in $q_S$ up to terms that are quadratic in $q$.
Notice that it vanishes rapidly in the limit $m_3/q_S\to 0$,
that is, relativistic motion for the spectator. 

 We can now define 
the velocity-ordered region as
\bea
\bar\beta_S < 1\, , \quad \frac{v}{\bar{\beta}_S} < 1\, ..
\label{vordered}
\eea
That is, we shall assume that it is possible to expand in $v/\bar\beta_S$,
keeping in mind that not all $v$ or $\bar\beta_S$ dependence 
is of this form.  Of course, such an expansion may be of limited 
quantitative use when
$\bar\beta_S \sim v$.  This is a region likely to be of particular
importance for charm quarks, where even $v$ need
not be small, and where the dynamics 
goes over into a true three-body problem.  
Nevertheless, we hope to gain insight 
even from the somewhat idealized kinematical
limit of Eq.\ (\ref{vordered}).

For small $\bar\beta_S$, Eq.\ (\ref{betaSdef})
shows that the vector $q_S$ is proportional to $\bar\beta_S$,
just as the active pair's relative momentum, $q$ is proportional to $v$.
In this way, we can rewrite  our result  in a form 
that exhibits the leading
 dependence on both velocities $v$ and  $\bar\beta_S$ in
 the velocity-ordered region,
\bea
\frac{1}{\sqrt{1-P_1^2P_3^2/(P_1\cdot P_3)^2}}
-
\frac{1}{\sqrt{1-P_2^2\,P_3^2/(P_2\cdot P_3)^2}}
 = -
 \frac{2}{\bar\beta^2_S}\ v\, \cos\phi_S\
\left[ 1 + {\cal O}\left(\frac{q^2}{m^2},
\frac{q\cdot q_Sq_S^2}{m\, m_3^3}\right)\right]\, ,
\nonumber\\
\label{betaminustwo}
\eea
where on the right $\phi_S$ is the
angle between $\vec q_S$ and $\vec q$ in
the rest frame of the active pair.
Equation (\ref{betaminustwo}) provides 
an explicit illustration of the enhancement
of gluon exchange as the spectator 
approaches the active pair in phase space.

We can summarize our analysis for the
imaginary NLO pole in the amplitude 
at small $q_S$ by using (\ref{betaminusthree}) and 
(\ref{betaminustwo})  in (\ref{allorder}), which gives
\ba
\sum_{i_1i_2}\, \delta_{i_1i_2}\ {\rm Im}
\left[{\cal A}_{13}+{\cal A}_{23}\right]_{i_1\dots i_4;i'_1\dots i'_4}
&\sim& -\, \frac{1}{\vep}\, \left(\frac{\alpha_s}{2}\right) \
\, \left(T_a\right)_{i'_2i'_1}\left(T_a\right)_{i_3i'_3}\ 
\delta_{i_4i'_4}\; \frac{2}{\bar\beta^2_S}\,
\left (1 - \bar\beta_S^2 \right )^{3/2}\, v\, \cos\phi_S\, .
\nonumber\\
\label{firstorder}
\ea
We will return to this expression in Sec.\ 5, when we discuss its possible
phenomenological implications for quarkonium production 
in association with open
heavy flavor.    In the remainder of this section,  we rederive the
basic result (\ref{firstorder}) by expanding first 
at the diagrammatic level.

\subsection{Operator interpretation and potential NRQCD}

We have observed that the eikonal factor, ${\cal A}$ of (\ref{allorder})
vanishes at $v=0$, that is, at vanishing active
pair relative momentum.  
The lowest order in $v$
exhibits an infrared singularity characteristic of
an electric dipole transition \cite{Nayak:2005rt},
although in this case the dipole is coupled to the
field of a spectator rather than to an emitted gluon.
The expansion
of Eq.\ (\ref{allorder}) shows this result at lowest order,
through its linear behavior on the relative momentum $q$.
In this section, however, we will expand the
diagrams first to order $q$, and sketch the resulting  
calculation. We may think of this as the 
first step in the creation of an effective theory
for color transfer, in the restricted region where
$v < \bar\beta(P_1,P_3),\bar\beta(P_2,P_3)$.

\begin{figure}[h]
\begin{center}
\begin{minipage}[c]{1.3in}
\epsfig{figure=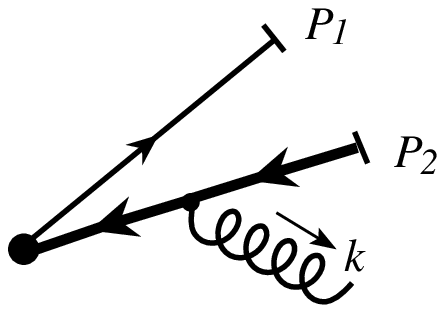,width=1.3in}
\end{minipage}
\hskip 0.2in
+
\begin{minipage}[c]{1.3in}
\epsfig{figure=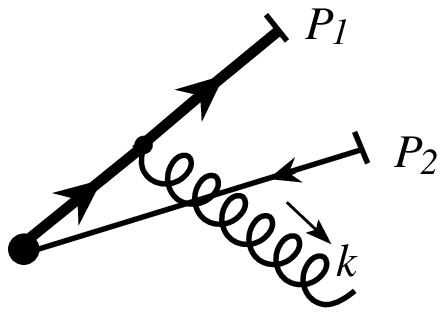,width=1.3in}
\end{minipage}
\hskip 0.2in
=
\hskip 0.2in
\begin{minipage}[c]{1.2in}
\epsfig{figure=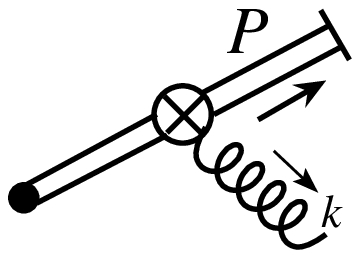,width=1.2in}
\end{minipage}

\caption{Graphical representation of the first-order expansion
in the relative velocity of the active pair.  The vertex $\otimes$
represents the inner product of the chromoelectric field
and the relative velocity, as in Eq.\ (\ref{Eadef}).
\label{fig5}}
\end{center}
\end{figure}

We represent the sum of the two eikonal
appoximations to the coupling of a soft gluon
to the active pair in the NLO eikonal factor
in  Fig.~\ref{fig5}.  As in Ref.\ \cite{Nayak:2005rt},
the double line on the right-hand side of the
figure represents the combined quark pair.  
As we shall see, the double line and
the vertex $\otimes$ corresponds to a 
Wilson line in adjoint representation that terminates 
at the field strength
operator $P_\mu F^{\mu\nu}q_\nu$, where $q$ is
the relative momentum of Eq.\ (\ref{Pqvdefs}).

Returning to the NLO integrals of Eq.\ (\ref{sf_nlo-b}), and 
recalling the relation
between $q^\mu$ and $v$ in Eq.\ (\ref{Pqvdefs}),
we see that an expansion in powers of $v$ is equivalent
to an expansion in $q$.  For this purpose, we
denote the spectator momentum by $l$, which 
may represent  $P_3$ or $P_4$ in this case. 
This was the notation used in Refs.\ \cite{Nayak:2005rt,Nayak:2006fm},
where, however, we took $l^2=0$.

In these terms, the order-$q$ correction to the two
contributions to the eikonal factor $\cal A$
of Eq.\ (\ref{factorme}) is given by 
\ba
{\cal A}_{13}^{(1)}+{\cal A}_{23}^{(1)}
= -i\, g^2\,\mu^{2\vep}
\int \frac{d^D k}{(2\pi)^D}\,
\frac{N(P,q,l,k)}
     {[P\cdot k + i\epsilon]^2\,
      [-l\cdot k + i\epsilon]\,
      [k^2 + i\epsilon]}\ +\ {\cal O}\left (\frac{q^2}{P^2} \right)\, ,
\label{sf_nlo-1}
\ea
where the expansion results in a squared denominator 
for $P\cdot k$ and in the numerator factor
\ba
N(P,q,l,k) =  
4\left[
q\cdot l \, P\cdot k - P\cdot l\, q\cdot k
\right]\, .
\label{numerator-nlo}
\ea
As shown in Ref.\ \cite{Nayak:2005rt}, 
we can  interpret this expression as
the local field-strength vertex appearing in
the nonlocal operator 
\ba
E_b \equiv
-ig\, \int_0^\infty d\lambda \, \lambda\
\left[\, P^\mu q^\nu F_{\nu\mu,a}(\lambda P)\, \right]
     \Phi_P^{(A)}(\lambda,0)_{ab}\, .
      \label{Eadef}
 \ea
In the rest frame of the active pair, this is precisely the integral
along a massive world-line, specified by momentum $P$, of
the scalar product of the relative momentum with
the electric field operator, contracted with
a gauge link in adjoint representation (defined
as in Eq.\ (\ref{complete})).  Notice the factor
$\lambda$ in the integral, which reflects the
increasing separation of the quark and antiquark
as the pair propagates into the final state at
fixed relative velocity $v=q/E^*$ (Eq. (\ref{Pqvdefs})).
The extra factor of $\lambda$ produces the squared denominator
in Eq.\ (\ref{sf_nlo-1}).  

In Eq.\ ({\ref{Eadef}), the connection of our NLO correction to 
the effective theory potential NRQCD (pNRQCD) \cite{Brambilla:1999xf}
is manifest.  The gauge link is the 
propagator for an octet pair in pNRQCD, and the field-strength 
is the operator in the Hamiltonian that transforms an
octet pair to a singlet configuration.   Although it was
not designed to describe quarkonium production,
pNRQCD operators emerge naturally
in the velocity-ordered region, as a description of
the active pair evolving in an external color field supplied
by the spectator.

We now turn to the evaluation of the NLO correction starting
from the effective vertex of (\ref{Eadef}).
It is straightforward to evaluate the integral in Eq.\ (\ref{sf_nlo-1}),
but we describe it in some detail because it sheds light on
the role of the spectator mass.
For simplicity, we may choose $l_\perp=0$ and $P=(M_H,\vec{0})$.
From Eq.~(\ref{numerator-nlo}), we then find, after a little algebra,
\bea
N = 4\,M_H\,q^3\, 
\left[ (l\cdot k) - 2l^+k^- \right]
+ 4P\cdot l\, q_\perp\cdot k_\perp\, ,
\eea
where the last term vanishes after integration over $k_\perp$.
The first term in square brackets, $l\cdot k$,
cancels the spectator denominator.  Because this term is actually
independent of the spectator momentum, it can be treated
in exactly the same manner as in the massless case
\cite{Nayak:2006fm}, leading to the same result.   
This contribution is purely real, and cancels against
real-gluon emission.  
We are left with the
contribution of the $l^+k^-$ term only, which we now
evaluate.

As in Ref.\ \cite{Nayak:2005rt}, we integrate over
$k^-$ first.   When we close the $k^-$ integration contour 
in the lower half-plane,
we encounter poles from both the $k^2$ and $P\cdot k$
denominators.   The $k^2$ pole gives a real result and
cancels against the corresponding numerator factor
for the final state with a gluon.
The remaining contribution in the $l^+k^-$ term in Eq.\ (\ref{sf_nlo-1})
is from the double pole at
$P\cdot k=0$ in Eq.~(\ref{sf_nlo-1}),
which gives
\bea
{\cal A}^{(1)}_{\ell^+}
&=& 
- \frac{16\, g^2\, \mu^{2\varepsilon}}{(2\pi)^{D-1}}
\left(\frac{q^3}{M_H}\right) 
\int_{-\infty}^\infty dk^+ \int d^{D-2}k_\perp\;
\nonumber\\
&\ &
\hskip 1.2in \times
\frac{d}{dk^-}\,
\left[\, \frac{l^+k^-}{(2k^+k^- - k_{2\perp}^2+i\ep)\, 
          (l^-k^+ + l^+k^- -i\ep)}\, \right]_{k^- = -k^+}
\nonumber\\
&=& 
\frac{16\, g^2\, \mu^{2\varepsilon}}{(2\pi)^{D-1}}
\left(\frac{q^3}{M_H}\right) 
\int_{-\infty}^\infty dk^+ \int d^{D-2}k_\perp\;
\frac{l^+}{[2(k^+)^2 + k_{\perp}^2 -i\ep]\, 
[\sqrt{2}l^3 k^+ +i\ep]}\, \
\nonumber\\
&\ &
\hskip 1.2in \times
\left[\,  \frac{2(k^+)^2}{2(k^+)^2+k_{\perp}^2} + 
\frac{l^-}{\sqrt{2}l^3}\,
\right]\, .
\label{sf-nlo-1n}
\eea
In the second equality we have taken the derivative 
with respect to $k^-$. 
In the resulting expression on the right, the first term
in brackets vanishes, because it is odd in
$k^+$, while the second term 
can be found from the pole at $k^+=0$, and has vanishing real part.
After integration over $k_\perp$, we obtain an imaginary piece for
the eikonal factor,
\ba
-i\,{\cal A}^{(1)}_{\ell^+} 
=
{\rm Im}
\left[\, {\cal A}_{13}^{(1)}+{\cal A}_{23}^{(1)}\, \right]
= 
\frac{1}{\vep}\, \left(2\alpha_s\right)\, 
\left(\frac{q^3}{M_H}\right) 
\left[\frac{l^2}{(l^3)^2}\right]\, ,
\label{imagnlopole-rest}
\ea
found here 
in the $P=(M_H,\vec{0})$ frame with $\ell_\perp=0$.
In a general Lorentz frame, the leading order eikonal factor is
\ba
{\rm Im}
\left[{\cal A}^{(1)}\right]
= -\; \frac{1}{\vep}\, \left(2\alpha_s\right) \,
\frac{P^2\, l^2}{[(P\cdot l)^2-P^2\,l^2]^{3/2}}\,
\left(l\cdot q\right)\, .
\label{imagnlopole}
\ea
This result, of course, agrees (up to color factors) with the
expansion of Eq.\ (\ref{allorder}) 
for the imaginary parts of the diagrams
in Fig.\ \ref{dashedgluon}.
As in Eq. (\ref{allorder}), the essential role
of the spectator mass, $\sqrt{l^2} \leftrightarrow \sqrt{P_3^2}$ is
manifest.  We also see that for
$P\cdot l\gg l^2$, the effect 
decreases as $(P\cdot l)^{-3/2}$.

Having arrived at a result identical to 
the expansion of Eq.\ (\ref{allorder}) at NLO,
we are now ready to use the operator
formalism to study color transfer
at NNLO in the velocity-ordered region.

\section{Color Transfer at NNLO in the Velocity-Ordered Region.}

So far, we have identified an IR pole in the NLO imaginary
part of the production amplitude for
a color singlet (active) quark pair, starting from an octet pair
at short distances, by using the eikonal approximation.
This contribution depends on the presence
of a massive source, the spectator, nearby in phase space.
The spectator catalyzes the octet-to-singlet transition of the pair,
in the sense that its own  color representation is unchanged.
The effect decreases 
rapidly as the relative velocity of the source
approaches unity, where the amplitude is readily matched to
NRQCD.  On the other hand, when the 
relative velocity between the pair and spectator is
itself nonrelativistic, the effect 
is potentially significant.    

Following the logic of Ref.\ \cite{Nayak:2005rt}, 
we explore the implications of these infrared
poles for the production rates of heavy quarkonia
in association with open heavy flavor.
We will work in the velocity-ordered region, (\ref{vordered}), where the 
active pair's relative velocity, $v$ is even smaller
than its velocity relative to the spectator.   We then expand in $v$. 

From the point of view of NRQCD, we will calculate soft gluon
corrections to the coefficient function 
$d\hat\sigma_{A+B\to  Q\bar{Q}[n]+X}(p_H)$
in Eq.\ (\ref{nrfact}), for the specific case of $Q\bar{Q}[n]$ a pair
of heavy quarks in singlet configuration.  We will identify
real infrared {\it double} poles in this production amplitude starting at 
two loops.
As above, at the level of
infrared poles in dimensional regularization, these
can be computed using eikonal approximation,
or equivalently, as vacuum expectations
values of products of Wilson lines, defined 
in adjoint and fundamental representations
as in Eq.\ (\ref{complete}).

\subsection{Matrix elements for pair production at NNLO}

To recall, Ref.\ \cite{Nayak:2005rt} dealt specifically with
infrared poles in fragmentation functions.
The relevant eikonal production fragmentation
function is given to lowest order in $q\sim mv$ by
\bea
    {\cal I}_2^{(8\to 1)}(P,q,\vep) &\equiv&
2\, \sum_N\, 
     \int_0^\infty d\lambda' \,  \lambda' \, \left< 0 \right| \,  
       \Phi_l^{(A)}{}^\dagger (\infty,0)_{bd'}\,
     \Phi_P^{(A)}(\lambda',0){}^\dagger_{d'a'}\,
        \left[\, P^\mu q^\nu F_{\nu\mu,a'}(\lambda' P)\, \right]\,
       \left|  N\right\rangle
       \nonumber\\
       &\ & \hspace{5mm} \times  \left\langle N\right|
    \int_0^\infty d\lambda \,  \lambda \,  \,
   \left[\, P^\mu q^\nu F_{\nu\mu,a}(\lambda P)\, \right]
     \Phi_P^{(A)}(\lambda,0)_{ad}\, 
     \Phi_l^{(A)} (\infty,0)_{db}\, |0\rangle\, ,
\label{I2def}
\eea
with (anti-) time ordering implicit in the (complex conjugate) amplitudes.
Here $\Phi_l^{(A)}$ is the same light-like Wilson line
in adjoint representation
link as in Eq.\ (\ref{complete}), which represents
the effect of recoiling massless quanta.  The other
Wilson line in adjoint representation, $\Phi_P^{(A)}$,
is  massive, and represents the propagation
of the pair of heavy quarks as an octet.   This
gauge link connects the hard scattering
at the origin with the field strength tensor
as in Eq.\ (\ref{Eadef}), which describes the absorption
of the soft gluon that changes the net color
of the pair from octet to singlet.  We note that
in this paper we choose the argument of the
field strength to be $\lambda'P$, rather
than $\lambda'P/2$, as in \cite{Nayak:2005rt}, which leads
to an explicit factor of 2 on the right-hand side of (\ref{I2def}).

In this section, we study an eikonal production cross
section, in which the light-like gauge link $\Phi_l^{(A)}$
of Eq.\ (\ref{I2def}) is replaced by a pair of massive
Wilson lines in quark and antiquark representation.
These two lines are linked together by an octet short-distance 
vertex (the shaded circles of the figures), 
which matches to the adjoint representation
of the line in the same direction as the total momentum, $P$, of the pair.
Again, the adjoint Wilson line terminates at 
the field strength vertex, Eq.\ (\ref{Eadef}),
\bea
    {\cal M}_2(P,q) &\equiv&
    \sum_N
     \int_0^\infty d\lambda' \,  \lambda' \, \left< 0 \right| \,  
     \Phi_{P_3}^{(q)}{}^\dagger (\infty,0)_{ij'}\,  
      \left(T_{c'}\right)_{j'k'}\, 
      \Phi_{P_4}^{(\bar{q})}{}^\dagger (\infty,0)_{k'l}\, 
     \Phi_P^{(A)}{}^\dagger(\lambda',0)_{c'a'}\,
        \left[\, P^\mu q^\nu F_{\nu\mu,a'}(\lambda' P)\, \right]\,
       \left|  N\right\rangle
       \nonumber\\
       &\ & \hspace{5mm} \times  \left\langle N\right|
    \int_0^\infty d\lambda' \,  \lambda' \,
   \left[\, P^\mu q^\nu F_{\nu\mu,a}(\lambda' P)\, \right]
     \Phi_P^{(A)}(\lambda',0)_{ac}\,   
     \Phi_{P_4}^{(\bar{q})} (\infty,0)_{lk}\,
      \left(T_c\right)_{kj}\, 
     \Phi_{P_3}^{(q)} (\infty,0)_{ji}|0\rangle\, ,
\label{M2def}
\eea
again with (anti-) time ordering implicit 
in the (complex conjugate) amplitudes.
This production cross section at NLO is represented 
by the diagram on the left in Fig.\ \ref{fig4}.
As in Fig.\ \ref{fig5}, the double line stands for the pair of 
heavy quarks at low relative velocity $v$, terminating 
in a singlet final state represented by the vertical line in the figure.
The vertex $\otimes$ supplies the necessary color.
We sum over all final states that include the singlet pair ($P$)
and the antiquark line $P_4$.

The diagram on the left of Fig.\ \ref{fig4} gives the entire NLO
contribution to ${\mathcal M}_2$, Eq. (\ref{M2def}).   It
represents the classic lowest-order octet mechanism,
in which the pair in octet representation emits a single 
gluon and becomes as singlet.   
It has a single infrared pole in dimensional regularization,
and is the lowest-order
contribution that is matched to nonperturbative matrix 
elements in NRQCD \cite{nloepsrefs}.   At this order, 
the associated pair is irrelevant to the infrared
structure, so long as we concentrate on
octet short-distance functions.   

The diagram of the same order on the right in Fig.\ \ref{fig4} represents an
interference between an octet pair produced at short distances
 in the amplitude with a singlet pair produced at short distances
 in the complex conjugate amplitude.  In the amplitude,
the pair is transformed into a singlet by the exchange of 
a soft gluon with the spectator quark line $P_3$.   We include this
interference diagram because it illustrates at NLO the cancellation
of infrared poles from an on-shell gluon ($k^2=0$) 
in virtual corrections and in the final state.   Both of
these states are available even when the heavy
active pair (double line in the figure) is fixed to be in
a singlet state.   After the cancellation of the real gluon contributions,
however, the imaginary pole that we studied in section 3 above
remains, and must be cancelled by complex conjugate diagrams
(not shown in the figure).

\begin{figure}[h]
\begin{center}
\epsfig{figure=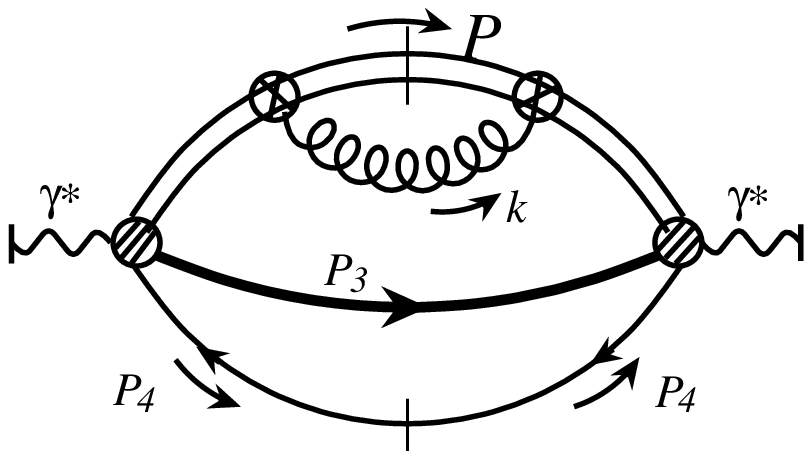,width=2.2in}
\hskip 0.5in
\epsfig{figure=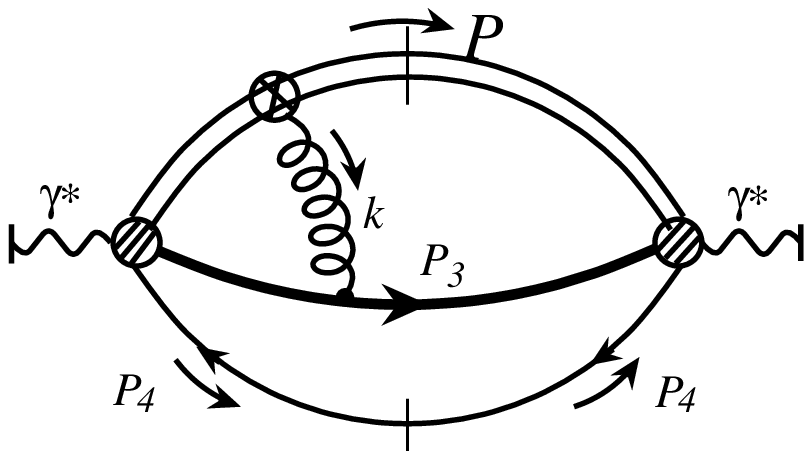,width=2.2in}
\caption{NLO contributions to $\gamma^* \to [Q\bar{Q}]Q\bar{Q}$ 
with the double line and the vertex $\otimes$ defined in
Fig.~\protect\ref{fig5}.  The cut double lines in each
case denote a heavy quark pair in a singlet state.  The shaded
circles of the diagrams represent short-distance functions
in the amplitude and complex conjugate.
In the figure on the left, both short-distance functions
are octet, corresponding to the matrix element ${\mathcal M}_2$
of Eq.\ (\ref{M2def}).   On the right, the short-distance 
function of the amplitude is octet, and of the complex
conjugate, singlet.   The right-hand figure, which illustrates
the cancellation of real and virtual on-shell gluons,
does not correspond directly to Eq.\ (\ref{M2def}).
\label{fig4} }
\end{center}
\end{figure}

Although the pole we have found above at one loop is purely
imaginary, it will contribute to the real part at NNLO.
This pole, of course, is just the one-loop contribution to a
Coulomb phase, and is guaranteed to cancel in a
fully inclusive cross section.  As we shall see, however,
when restrictions are placed on the color of final-state
pairs, as in NRQCD, this cancellation fails in general at NNLO.

\subsection{The analysis of double poles at NNLO}

The NNLO diagrams that contribute to the squared octet-to-singlet
transition probability, Eq.\ (\ref{M2def}) are shown 
in Fig.\ \ref{softgluonfig}.
In these diagrams, the cut double line represents the active pair 
in color singlet configuration, and the crossed circles the field-strength
vertices.
All  lines in these diagrams except for gluons are eikonal.
The transition probability  (\ref{M2def}) has been constructed 
to reproduce the infrared
behavior of diagrams in full QCD at leading order in the active-pair 
relative velocity, $v$.

The graphs shown in Fig.\ \ref{softgluonfig}
each represent a set of cut diagrams, found by summing over 
all final states that include the color singlet pair and the line $P_4$.
As in Fig.\ \ref{fig4}, the shaded circles represent short-distance
functions, which link the active pair in octet representation with the 
associated pair ($P_3$ and $P_4$).   
We work in the velocity-ordered region, Eq.\ (\ref{vordered}), 
and take the line labeled $P_3$
as the spectator that interacts with the pair.  We will neglect
connections of the active pair with the remaining line,
 $P_4$.   That is, we assume that
the relative velocity of this line to the pair is large. 

Our goal in this section is to show that at NNLO
there is a real infrared-sensitive contribution to
the octet-to-singlet transition amplitude, ${\mathcal M}_2$.
We will not calculate the diagrams explicitly,
but argue that in dimensional regularization 
diagram (I) produces a double pole that equals
the absolute square of the NLO result, Eq.\ (\ref{firstorder}) 
(up to color factors),
and that there is no other source of real, double
infrared poles from any other diagram. 
Such a real contribution to the transition amplitude cannot
be matched to a local operator in NRQCD.
We will rely heavily on power counting analysis,
and on the specific calculations of Ref.\ \cite{Nayak:2005rt}.

\begin{figure}[h]
\begin{center}
\epsfig{figure=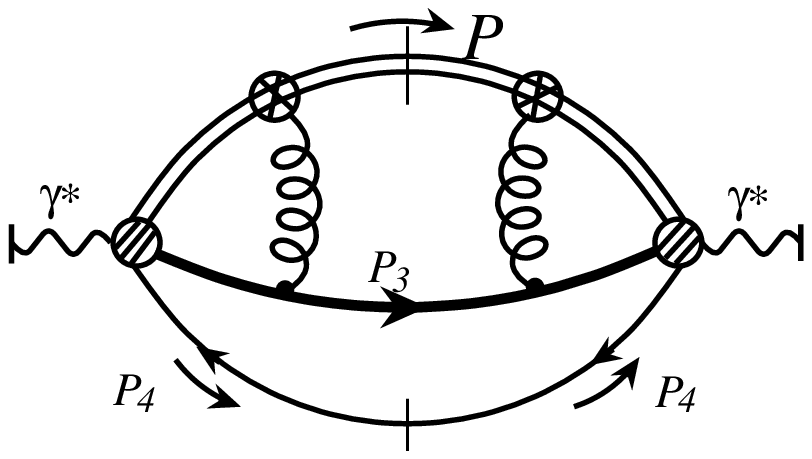,width=2.0in}
\hskip 0.7in
\epsfig{figure=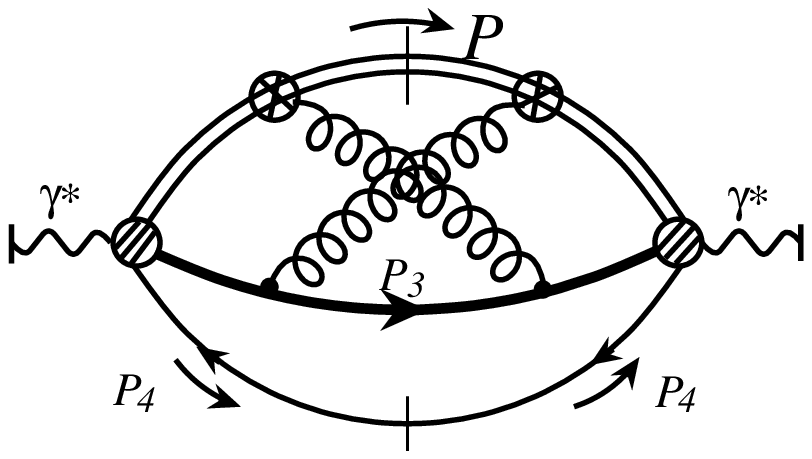,width=2.0in}

(I) \hskip 2.5in
(II) 
\vskip 0.2in

\hskip 0.6in
\epsfig{figure=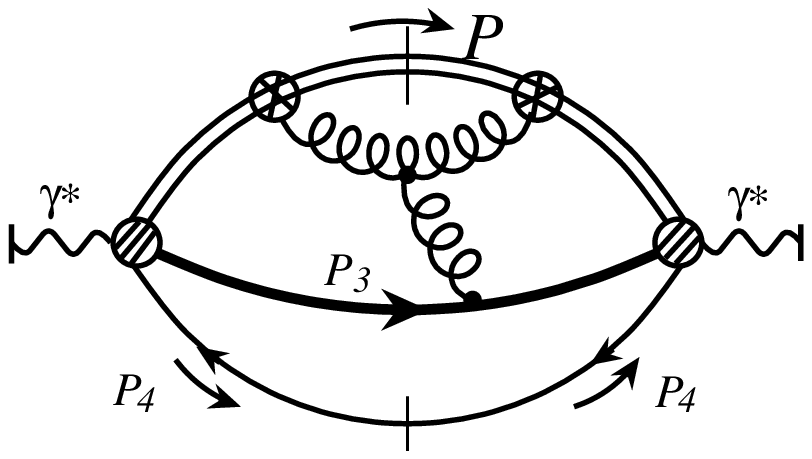,width=2.0in}
\hskip 0.8in
\epsfig{figure=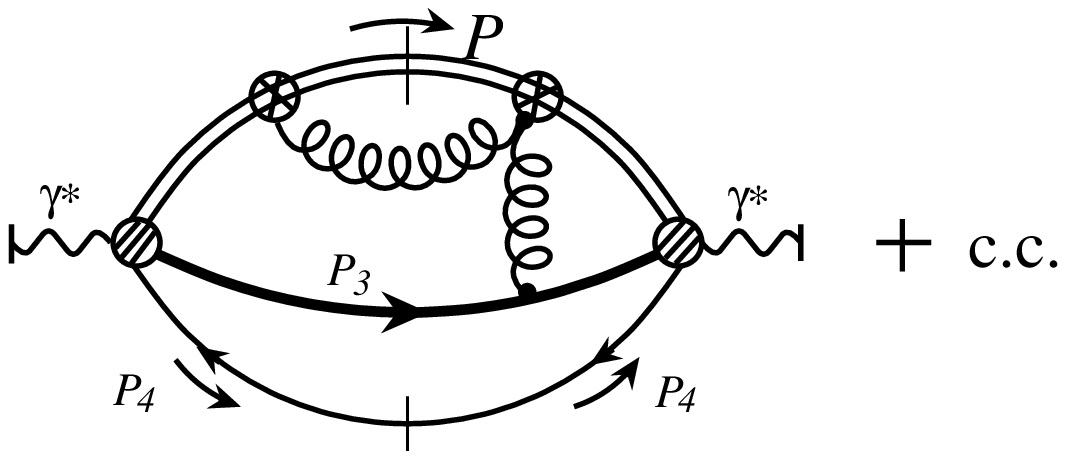,width=2.6in}

(III)\hskip 2.5in
(IV) 
\vskip 0.2in

\hskip 0.6in
\epsfig{figure=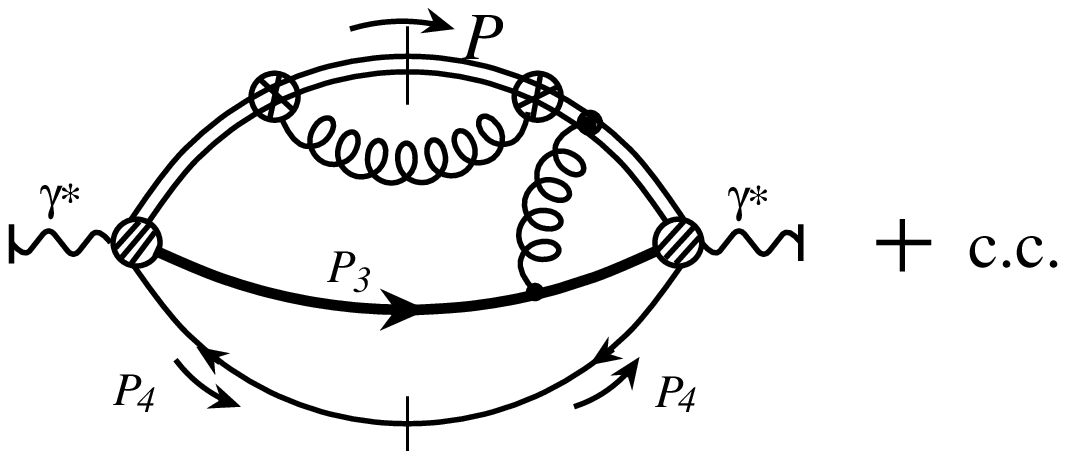,width=2.6in}
\hskip 0.2in
\epsfig{figure=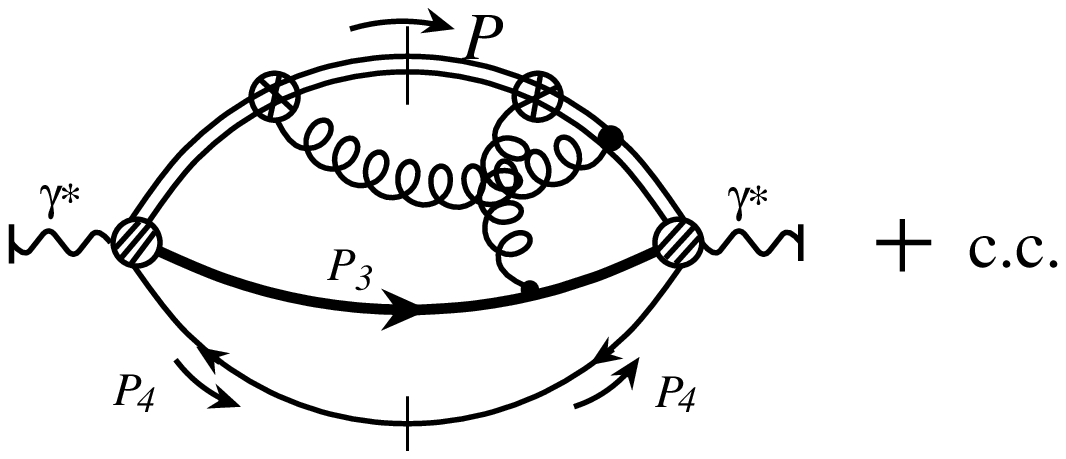,width=2.6in}

(V) \hskip 2.5in
(VI)

\caption{Diagrams that contribute to ${\cal M}_2$,
Eq.\ (\ref{M2def}), similar to those studied in Ref.\ \cite{Nayak:2005rt}.
These NNLO diagrams describe the transition of a pair
of momentum $P$ from octet to singlet by the exchange of color 
with a spectator whose momentum, $P_3$ is nearby in phase space.
We sum over all cuts of these diagrams that can
produce a color singlet quark pair.   
\label{softgluonfig}}
\end{center}
\end{figure}

To generate a double pole in $D=2-\varepsilon/2$ dimensions, 
a diagram must contain two momentum integrals, either virtual
or real, that diverge logarithmically when $D=4$.
This can be checked by straightforward 
power counting.   General rules for power counting estimates
of infrared-sensitive diagrams
are discussed, for example, in Ref.\ \cite{Sterman95}.
The prescription is particularly simple for massive eikonal lines
exchanging gluons without gluon self-coupling vertices.
This describes all the diagrams of Fig.\ \ref{softgluonfig} 
except for (III).   Because all the
eikonal lines are massive, there are no collinear singularities
anywhere in loop momentum or
phase space for diagrams (I), (II) and (IV)-(VI).
In contrast, we do find collinear singularities for some momentum
configurations of diagram (III) because of the three-gluon coupling
in that diagram.
These collinear singularities, however, cancel
in the sum over cuts at fixed active pair color
by standard unitarity arguments \cite{Sterman95},
and we need not  consider them here.

To perform power counting for the remaining, infrared singularities, we
need to study all choices of loop momenta carried by the gluons.
The momenta of the active pair is fixed, as are $P_3$ and $P_4$.
When a gluon appears in the final state, we do power counting
on its  phase space integral, which we shall
continue to refer to as a ``loop" momentum.   
From a technical point of view, in
the definition of ${\mathcal M}_2$, the vertices associated with
the short-distance functions are fixed at the origin in both
amplitude and complex conjugate.   These vertices are
therefore relatively local, and a loop momentum can begin at
the short-distance function in the amplitude on the left of a diagram,
and end at the short-distance function on the right, in the complex
conjugate amplitude.   In physical terms, because we neglect
the recoil of the heavy quarks, we do not impose overall
momentum conservation on soft gluons in the final state.

We scale the components of any such choice of loop momenta as
\bea
k_i^\mu = \lambda_i \hat  k_i^\mu\, ,
\label{scale}
\eea
where $i=1,2$ labels a choice of loop momenta.
This is just a change of variables, which will
produce an overall factor $\lambda_i^4$ for each loop.
When the integrand behaves as $\lambda_i^{-p_i}$
with $p_i=4$, all
sizes of loop momenta contribute equally to
the result, and the integral diverges logarithmically.
In fact, because these eikonal integrals are
scaleless overall, there is always a single
logarithmic divergence in both infrared and
ultraviolet when both loops are scaled together, 
and the diagrams are nonzero
in dimensional regularization only after imposing an ultraviolet cutoff.

The application of the scaling (\ref{scale}) is
quite straightforward, keeping in mind several technical
observations.

\begin{enumerate}

\item[1.] To derive a double infrared pole, it is necessary to
find logarithmic ($p_i=4$) power counting in both the $\lambda_i$,
$i=1,2$, individually.   
When any virtual eikonal line carries both loop momenta, we can
neglect the dependence of that line on the softer momentum.
We can thus do power counting in
the softer loop alone, at fixed values
of the larger loop momentum.  If $p_{\mathrm soft}<4$
there is at most a single pole.   On the other hand,
if we find any one-loop {\it subdiagram} with logarithmic power
counting for a given final state, dimensional analysis ensures
that there is a double pole.

\item[2.] The virtual spectator eikonal lines in the $P_3$ direction
(middle lines of the diagrams),
contribute $\lambda_i^{-1}$, with $i$ corresponding
to the larger of the loop momenta that flow 
through the line.  

\item[3.]  When there is a single active pair eikonal line 
at the top of a diagram
in the amplitude or complex conjugate,
it gives $\lambda_i^{-2}$, with $i$ the larger
loop momentum carried by the line.  The extra factor of $\lambda_i^{-1}$
relative to the normal eikonal line
corresponds to a modification of the standard eikonal
Feynman rules \cite{Nayak:2005rt} due to 
the integral that defines the adjoint eikonal (double)
line ending in the field strength vertex, Eq.\ (\ref{M2def}).
We have encountered this effect above in the one-loop calculation, 
Eq.\ (\ref{sf_nlo-1}).
The three-point crossed vertices, however,
 each contribute a 
numerator factor of $\lambda_i$ because of the linear dependence on
loop momentum in their definition, and four-point crossed
diagrams are independent of loop momenta.
The net effect of the combination of 
a field strength vertex and a single active pair eikonal
line is an overall $\lambda_i^{-1}$
when a gluon of momentum $k_i$ attaches to
the vertex.    

\item[4.] When there are two  active pair eikonal lines,
as in diagram (V), the diagram represents the sum of two
terms, in which one of the two eikonal denominators is squared.  
For example, taking $k_1$ as the gluon connecting the field
strength vertices, and $k_2$ the momentum of the
gluon flowing (up) from $P_3$ to the active eikonal in diagram (V),
the two active eikonals on the right correspond to the combination
\bea
\frac{1}{(P\cdot k_1)^2}\; \frac{1}{P\cdot (k_1+k_2)}\ + \ 
\frac{1}{P\cdot k_1}\; \frac{1}{(P\cdot (k_1+k_2))^2}\, .
\eea

\end{enumerate}

We are now ready to discuss the diagrams of Fig.\ \ref{softgluonfig} 
individually.

{\it Diagram (IV)}:
Let us begin the application of the above rules with diagram (IV)
 of Fig.\ \ref{softgluonfig}, and show that it
can generate at most a single infrared pole.
We must test all independent choices of soft
gluon loops.   To derive a double pole, we must find one loop
for which scaling according to Eq.\ (\ref{scale})
at fixed values of the other loop's momentum
results in a power $\lambda_i^{-4}$.

Again, let $k_1$ be the momentum of the gluon
that connect the two crossed vertices,
and suppose $k_1$ is the larger loop momentum.
It must then flow out from the short-distance vertex
on the left, and may flow to the short-distance vertex on the right
either along the double line at the top, 
or down to the $P_3$ line along the
vertical gluon of the figure, and from
there to the right-hand hard vertex.

From our observations above, we need only do power
counting for the remaining loop, $k_2$, to check for double poles.
If the vertical gluon is virtual, the $k_2$ loop flows
through the right-hand vertex
and the vertical gluon line.  Power
counting for this loop depends on how the
$k_1$ loop momentum flows.
If $k_1$ flows along the top double line,
we find $p_2=3$, while if $k_1$ flows
down to $P_3$ along the virtual gluon,
$p_2=2$. In 
both cases $p_2<4$, and the $k_2$
loop cannot produce an independent logarithmic divergence.

The power counting is a bit different 
if $k_2$ describes a gluon in the final state.
Since $k_2$ is the softer gluon, $k_1$ cannot
flow (backwards) through the $k_2$ gluon,
but must flow only along the double line
on top.   The power counting for the softer line then again gives
$p_2=2$.   

Finally, if $k_2$ defines the harder of the two gluon momenta,
we readily verify that the $k_1$ loop is infrared finite.
In summary, Diagram (IV) cannot
produce a double pole in dimensional regularization.

In fact, diagram (IV) is the only diagram in Fig.\ \ref{softgluonfig}
whose individual cuts lack a double pole.   We now show, however, that 
for all of the
remaining diagrams except for (I), the real parts of the residues of
such double poles cancel in the sum over 
allowed cuts (final states) for each diagram.
Imaginary double poles, of course, cancel
when diagrams are combined with their conjugates.
We will find a real double infrared pole from diagram (I).

{\it Diagram (V)}:  Let $k_1$ be the momentum of the gluon connecting
the field strength vertices,  and $k_2$ the momentum
of the gluon exchanged between the active pair and
spectator $P_3$.   From the power counting rules
described above, it is clear that a double pole is
produced by taking $k_1$ as the softer loop momentum.
There are two final states to consider, one for which the gluon
$k_2$ is virtual, and  the other in which
it is ``real", that is, it appears in the final state.  

Suppose that gluon  $k_2$ is virtual.
Taking the $\lambda_1\to 0$ limit, the $k_2$ integral
becomes the same NLO expression as in Eq.\ (\ref{nlointegral}) above
(for diagram (V) it is actually the complex conjugate).
As in that case, there are two contributions,
associated with poles in the complex $k_2^-$
plane at $k_2^2=0$ and
$P\cdot k_2=0$.   The former is purely real,
the latter purely imaginary.  When combined
with the $k_1$ integral, they both produce $1/\varepsilon^2$
contributions in dimensional regularization.
The imaginary pole, of course, cancels against
the complex conjugate diagram.  
This leaves the real double pole, associated
with an on-shell gluon $k_2$.   This
singularity cancels against the final state
in which the gluon $k_2$ is real, point-by-point in the remaining
momentum integrals \cite{Nayak:2005rt}.

{\it Diagram (VI)}:  
In diagram (VI) double poles arise
from a momentum configuration that
is complementary to that of diagram
(V).   By analogy to the latter,  we choose $k_2$ as
the gluon attaching the active pair
with $P_3$, and $k_1$ the remaining
gluon, emitted from the field strength vertex on
the left of the diagram.   The double pole then
arises by taking $k_2$ as the softer loop
momentum.   At fixed values of $k_1$,
however, the infrared poles arising from
the $k_2$ integral cancel just as in the
NLO case.  Again, no double pole
survives when diagram (VI) is added
to its complex conjugate.

{\it Diagram (III)}:
In diagram (III),
there are actually two relevant loop assignments
for which the softer loop has logarithmic power counting.
To describe them, we denote the momentum flowing
out of the field strength vertex on the left as $k_1$,
and the momentum flowing into the field strength vertex
on the right as $k_2$. The momentum of the gluon
attached to the $P_3$ (middle) line
is then $k_2-k_1$ (flowing up).
The double logarithmic scalings are then
$\lambda_1 \gg \lambda_2$ and
$\lambda_1 \sim \lambda_2 \gg \lambda_2-\lambda_1$.
In the latter case, the larger momentum flows 
out of the active pair eikonal and back, and in the
former case, it flows down to the $P_3$ eikonal line.
Both cases correspond to double 
logarithmic power counting, and in the 
calculation of Ref.\ \cite{Nayak:2005rt}, both
regions contributed to a real single pole.

The calculation  of (III) was described
in great detail in Ref.\ \cite{Nayak:2005rt},
and the leading, double poles were found
to follow the same pattern as for
diagrams (V) and (VI) above.  That is,
real double poles cancel between final states
in  which  the gluon that attaches to the
$P_3$ line is either real or virtual, leaving
purely imaginary double poles only.
The loop integrations that led to this
result are the same as in our case,
except that the spectator was taken
massless, $P_3^2=0$.   Relaxing this
condition, it is straightforward to 
verify that the calculations follow
exactly the same pattern as in \cite{Nayak:2005rt},
and that diagram (III) does not contribute
a double pole after a sum over final  states.
It is worth noting that most
of the complications of \cite{Nayak:2005rt} 
result from going to the level of the subleading, real single pole.

{\it Diagram (II)}:
For diagram (II),  the pattern is the same as for 
(V) and (VI): the real double poles are
associated with on-shell gluons, which cancel,
leaving only an imaginary double pole proportional
to the result found at NLO in  Eq.\ (\ref{imagnlopole}).  
The pattern is again exactly as in the
case of a massless eikonal spectator \cite{Nayak:2005rt}.

{\it Diagram (I)}: 
We are left with diagram (I)
as the only remaining  potential source of a real double pole.
Diagram (I) is associated with final states that include
one, two or no gluons.   Exactly as in the massless
case \cite{Nayak:2005rt}, the final states with
one or two gluons cancel against the corresponding
contributions from NLO virual diagrams with one or
two gluon lines on-shell.   (Notice that two-loop virtual
corrections do not contribute in the octet-to-singlet
transition amplitude,  ${\mathcal M}_2$.)
This real/virtual cancellation, however,  leaves
the absolute  square of the imaginary
single-pole term of Eq.\ (\ref{firstorder}), identified in
Fig.\ \ref{fig4}.  Again, this contribution is proportional  to $P_3^2$,
and was hence absent in the massless case studied
in \cite{Nayak:2005rt}.   The complete infrared sensitive
result thus takes the form
\bea
{\cal M}_2 \sim \frac{\alpha_s^2}{\varepsilon^2}\  v^2\,
\ \frac{(1-\bar\beta_S^2)^3}{\bar\beta_S^4}\, ,
\label{M2sim}
\eea
which exhibits a strong dependence on relative veloctiy.

We should, of course, emphasize that
we have not evaluated the single poles
in these diagrams.  For massless spectators,
these poles can be factorized as in Ref.\ \cite{Nayak:2005rt},
but in this case we may safely assume that like
the double poles they will depend on the kinematics of heavy particles in the
final state.   In any case, having found uncancelled
double poles, we have already
demonstrated the infrared sensitivity 
of color transfer, independent of the
structure of the single poles.
In the following section, we discuss possible physical
implications of these uncancelled poles and the
consequent infrared sensitivity to color transfer
in associated production.

\section{Color Transfer in Heavy Quarkonium Production}

We have shown that the 
imaginary $1/\varepsilon$ NLO corrections in Eqs.\ (\ref{firstorder}) 
and (\ref{imagnlopole})
 result in real contributions to the
octet-to-singlet transition amplitude, ${\mathcal M}_2$, at NNLO.
As illustrated in Eq.\ (\ref{M2sim}) these corrections are 
conveniently expressed in terms of $\bar\beta_S$,  Eq.\ 
(\ref{betaSdef}), which measures the relative velocity
of the spectator and the pair, as in Eq.\ (\ref{betaminusthree}).  
We are
naturally led to suggest that 
these infrared sensitive corrections
tend to increase the cross section 
for bound states, since we have found a new
source of pairs with singlet
color  at small $\bar\beta_S$.
This
does not yet give us an estimate for its
effect on the production cross sections for
quarkonia.   To make such an estimate, we rely on an analogy
to the physical picture at the basis of the
color octet mechanism as it appears in NRQCD.

\subsection{Estimating color transfer}

Consider first the standard color octet mechanism.
In this case, a (real) infrared pole is associated with the
transition from a color octet to color singlet 
pair at NLO through an electric dipole
transition \cite{bodwin94}.   Each such  transition is associated
with a factor of the active pair relative momentum, $v$
in the amplitude, and $v^2$ in the cross section.
We can summarize this perturbative result as 
\cite{bodwin03,nloepsrefs}
\bea
d\sigma_{\rm octet}^{\rm (PT)} \sim
 d\hat\sigma_{e^+e^-\to  Q\bar{Q} [S_8]}(p_H)\
\frac{1}{\varepsilon}\ \frac{\alpha_s}{\pi}
v^2\, ,
\label{octetir}
\eea
with $p_H$ the momentum of the heavy quarkonium,
which is identified with the momentum of the active pair.
In NRQCD, such infrared poles are
matched with (equivalently, factorized into) color octet
matrix elements, $\langle {\cal L}_8^H\rangle$,
for the final-state  quarkonium, $H$.
   In this notation, ${\cal L}$ 
denotes the orbital angular momentum  
of the relevant operator.
For $J/\psi$ and similar quarkonia, 
$L=0$ operators give the largest contributions,
which requires at least two electric dipole
transitions. This produces a greater
suppression of $v^4$ in the cross section,
and hence the corresponding matrix element, 
$\langle{}^3{\mathcal S}_8^H \rangle$.
We thus make the replacement
\bea
  d\hat\sigma_{e^+e^-\to  Q\bar{Q} [S_8]}(p_H)\
\frac{1}{\varepsilon}\ \frac{\alpha_s}{\pi}
v^2
\ \rightarrow \
d\hat\sigma_{e^+e^-\to  Q\bar{Q}
[S_8]}(p_H)\, 
 \langle{}^3{\mathcal S}_8^H \rangle\, ,
\label{octetreplace}
\eea
in which the infrared-sensitive correction at NLO is 
matched to the nonperturbative $S$-wave matrix element,
even though the latter has different scaling in $v$.

Now consider the NNLO color transfer cross section in
associated production, including the
square of the full $\bar\beta$-dependence at lowest order in $v$,
given  in Eq.\ (\ref{betaminusthree}).  
Corresponding to Eq.\ (\ref{octetir}), we have
\bea
d\sigma_{\rm transfer}^{\rm (PT)} \sim
 d\hat\sigma_{e^+e^-\to  Q\bar{Q}
[S_8]+Q'(\beta_S)}(p_H)\
\frac{1}{\varepsilon^2}\ \alpha_s^2\, v^2\,
\ \frac{(1-\bar\beta_S^2)^3}{\bar\beta_S^4}\, ..
\label{transfer_pt}
\eea
The kinematic enhancement in this cross section associated
with small $\bar\beta_S$ is very strong whenever
$\bar\beta \le \sqrt{v}$, corresponding to low
relative spectator-active velocities.   
The quadratic $v$-dependence here is just the square of the
linear  $v$ in Eq.\ (\ref{betaminustwo}), which is
 the same as that of an electric dipole
transition.   As for the color octet mechanism, however, 
we anticipate that 
two  transitions proportional to the
dipole moment (and hence to $v$) will 
be necessary to produce a color singlet, S-wave quarkonium.
In the associated production cross sections
at hand, the anticipated $v^4$ suppression 
that is normally absorbed into the matrix element
is compensated, at least in part, by the explicit
factor $1/\bar\beta_S^4$.     In effect, for
color transfer the expansion in $v$
alone is replaced by an expansion in $v/\bar\beta$.

With our analogy in mind, it seems natural 
to replace the factor $\alpha_s^2\, v^2/\varepsilon^2$ 
in (\ref{transfer_pt})
by the same octet matrix element as in (\ref{octetreplace}) . 
In making such a replacement, we are assuming that
for color transfer
it costs the same overall factor of $v^4$ as in  the color octet
mechanism to
produce an S-wave, color singlet state for the active pair,
and also that 
once the active pair is in a color singlet state,
it evolves independently of the spectator.
We thus summarize our estimate of the full cross section 
for color transfer by
\bea
d\sigma^{\rm tot}_{e^+e^-\to H+X}(p_H) 
&\sim& d\hat\sigma_{e^+e^-\to 
Q\bar{Q}
[S_1]+Q'(\beta_S)}(p_H)\, 
 \langle{}^3{\mathcal S}_1^H \rangle
 \nonumber\\
 &\ & 
\hspace{10mm} +\
 d\hat\sigma_{e^+e^-\to  Q\bar{Q}
[S_8]+Q'(\beta_S)}(p_H)\, 
\frac{ \langle{}^3{\mathcal S}_8^H \rangle}{\bar\beta_S^4}\, 
\left (1 - \bar\beta_S^2 \right )^3\, .
\label{colortransfersigma}
\eea
We emphasize that our arguments give this expression at best
a heuristic justification, although we consider it a conservative estimate
of the color transfer contribution to associated production.
We do not rule out the possibility that yet higher orders might lead to 
greater enhancement in $\bar\beta_S$, but we leave this
to further investigation.
Another approach to estimate color transfer
in the velocity-ordered region is based
on the observation that the active pair is not 
truly on-shell, but should be treated according
to pNRQCD methods \cite{Brambilla:2004jw,Kniehl:2002br}.  
The transverse momentum
integrals that result in $1/\varepsilon$ poles 
in Eq.~(\ref{nlointegral}), for example, would then
be replaced by logarithms of the bound-state
relative velocity. 
In any case, to get a better idea of the role
of color transfer, we should compare the sizes 
of the individual singlet and octet
hard-scattering functions in Eq.\ (\ref{colortransfersigma}), as they
appear in the three-particle phase space of the heavy quarkonium and
the associated pair.   We turn now to this disucssion.

\subsection{Comparison of octet and singlet hard scattering}

The color transfer mechanism is most relevant to the 
kinematic region where the active heavy quark pair is 
in close vicinity with another heavy (anti)quark.  
As observed above, color transfer
could significantly affect the 
rate of heavy quarkonium associated production, such as 
$e^+ e^- \rightarrow J/\psi + c\bar{c}+X$, when the 
collision energy $\sqrt{s}$ is not too much higher than 
the mass threshold, $4m$, and the cross section is 
dominated by the region of phase space where one of 
the spectator heavy quarks is close to the active heavy 
quark pair.  
In addition,  the mechanism should
also have a strong effect 
on the size of the input distribution of heavy quark
fragmentation functions, e.g., 
$D_{c\rightarrow J/\psi}(z,\mu_0,m)$,
when the fragmentation scale $\mu_0$ is close to 
$3m$.  In either case, the radiation of light quanta will
tend to decrease the effective $s$, and make
it easier to produce spectators at low $\bar\beta$.

At collision energy $\sqrt{s}=10.6$~GeV, as
seen by the BELLE and BABAR collaborations, 
both heavy spectators in $J/\psi$ associated 
production can be highly
relativistic if the two pairs carry all the energy.
We will find evidence in this subsection 
that even in this case the influence of the color transfer mechanism on
$J/\psi$ production in association with $c\bar{c}$ 
could still be significant.  

From Eq.~(\ref{nrfact}), the lowest order
 inclusive prompt 
production of a heavy quarkonium with an additional 
$Q\bar{Q}$ in NRQCD
 is  given by
\begin{eqnarray}
\sigma^{\rm LO}_{e^+e^-\to H+Q\bar{Q}}(s) 
&=& 
\hat\sigma_{e^+e^-\to Q\bar{Q}[S_1]
                      +Q'\bar{Q}'}(s)\, 
 \langle{}^3{\mathcal S}_1^H \rangle
 \nonumber\\
&\ & 
+\
\hat\sigma_{e^+e^-\to  Q\bar{Q}[S_8]
                      +Q'\bar{Q}'}(s)\, 
 \langle{}^3{\mathcal S}_8^H \rangle\ .
\label{lowestorder}
\end{eqnarray}
Although the direct octet contribution from
${}^3{\mathcal S}$ to the $J/\psi$ production 
rate at $\sqrt{s}=10.6$~GeV is only about three percent of 
the singlet contribution \cite{Liu:2003jj},
the production rate, or coefficient function,
for the active heavy quark pair is actually much higher for
the octet mode than for the singlet, 
$\hat\sigma_{e^+e^-\to  c\bar{c}[S_8]+c'\bar{c}'}(s)
\gg \hat\sigma_{e^+e^-\to c\bar{c}[S_1]+c'\bar{c}'}(s)$.
The octet mode is suppressed because
the octet NRQCD matrix element
$\langle{}^3{\mathcal S}_8^{J/\psi} \rangle$ is much smaller
than the singlet matrix element 
$\langle{}^3{\mathcal S}_1^{J/\psi} \rangle$, by about two 
orders of magnitude \cite{Brambilla:2004wf}.  
In the estimate we have given above,
Eq.~(\ref{colortransfersigma}), however,
the contribution via the 
color transfer mechanism could be very significant if the 
enhancement factor in $\bar{\beta}_S$ 
compensates for the suppression of the octet matrix
element relative to singlet.   
 
The inclusive rate of charmonium production associated with 
an additional pair of heavy quarks in $e^+e^-$ annihilation
has been studied extensively in the NRQCD formalism 
at both leading order (LO) \cite{Liu:2003jj,inclusive-lo} 
and next-to-leading order (NLO) \cite{Zhang:2006ay}.
The LO analytic expression for the singlet production 
of various charmonium states in terms of the variable 
$z=2E_{c\bar{c}}/\sqrt{s}$ is available 
\cite{Liu:2003jj,inclusive-lo},
while numerical results for the inclusive rate are
available for the NLO contribution \cite{Zhang:2006ay}.
To better understand the contributions to the inclusive rate
of prompt $J/\psi$ production from different parts of phase space, 
we express the rate in terms of a phase space integration over
the invariant masses of the active heavy quark pair and the
heavy spectator (anti)quark:
\begin{equation}
s_3 \equiv (P+P_3)^2, \quad
s_4 \equiv (P+P_4)^2\, ,
\end{equation}
with $P^2=(2m)^2$ and $P_3^2=P_4^2=m^2$.
We denote the total squared collision energy by $s=(P+P_3+P_4)^2$.
The LO perturbative coefficient functions 
(or at this order, the partonic cross sections to
produce the state $Q\bar{Q}[S_n]$) in Eq.~(\ref{lowestorder})
is given by
\begin{eqnarray}
d\hat\sigma_{e^+e^-\rightarrow Q\bar{Q}[S_n]+Q'\bar{Q}'}(s) 
&=&\frac{1}{2s}\, \frac{1}{3}\, \frac{1}{c_n}\,
\overline{\sum}_\lambda
\left|{\rm Tr}[
 {\cal A}_{e^+e^-\rightarrow Q^i(P/2)\bar{Q}^j(P/2)+Q'\bar{Q}'}
 {\cal P}_{1\mu}]\,
 \epsilon^\mu_\lambda(P) \langle 3i\bar{3}j|n\rangle
\right|^2 
\nonumber \\
&\ & 
\times
\frac{d^3P}{(2\pi)^3 2E_{Q\bar{Q}}}
\frac{d^3P_3}{(2\pi)^3 2E_{Q'}}
\frac{d^3P_4}{(2\pi)^3 2E_{\bar{Q}'}}
(2\pi)^4\delta^4(P_{e^+}+P_{e^-}-P-P_3-P_4)
\nonumber\\
&=& 
\sigma_0\, e_c^2\, \frac{\alpha_s(s)^2}{12s^2}
\left(-g_{\mu\nu}W^{\mu\nu}_{(n)}(s,s_3,s_4,m) \right)\, 
\theta\left(\Phi(s,s_3,s_4,m)\right)\, 
ds_3\, ds_4\, , 
\label{lowestsigmahat}
\end{eqnarray}
where $n=1,8$, $c_n = 2N_c,N_c^2-1$, and 
$\langle 3i\bar{3}j|n\rangle =
 \delta_{ij}/\sqrt{N_c},\sqrt{2}(T^a)_{ji}$
for the singlet and octet contributions, respectively.
The overall factor $1/3 = 1/(2J+1)$ reflects $J=1$, 
the spin of the active $Q\bar{Q}$ pair.  In the trace, the projection
operator for a spin-1 $Q\bar{Q}$ pair is 
${\cal P}_{1}^\mu=(\gamma\cdot P/2 - m)\gamma^\mu 
                  (\gamma\cdot P/2 + m)/\sqrt{8m^3}$
\cite{Berger:1980ni}.
In the second equation in Eq.~(\ref{lowestsigmahat}),
$\sigma_0=4\pi\alpha_{em}^2/3s$, and $e_c=2/3$ is the
charm quark fractional
charge.    Finally, explicit forms for the hadronic tensor $W^{\mu\nu}_{(n)}$
with $n=1,8$ and the phase space constraint 
$\Phi(s,s_3,s_4,m)$ are given in the Appendix.

In Fig.~\ref{loxsec}, we show the integrand of the $s_3\, s_4$
integration for the production rate 
$\hat\sigma_{e^+e^-\rightarrow Q\bar{Q}(n)+Q'\bar{Q}'}(s)$ 
defined in Eq.~(\ref{lowestsigmahat}) 
with an active heavy quark pair in a singlet (left) and octet (right) 
color state.  To generate the figures in Fig.~\ref{loxsec},
we used $\sqrt{s}=10.6$~GeV, $m=1.5$~GeV.  
The units for the vertial axes are pb/GeV$^3$.
After integrating over the phase space of $s_3$ and $s_4$,
multiplying the same singlet NRQCD matrix element to J/$\psi$,
$\langle{}^3S_1\rangle = 
 \langle {\cal O}^{{\rm J/}\psi}_1({}^3_1)\rangle = 1.16$~GeV$^3$,
and using the same value of $\alpha_s$
used in Ref.~\cite{Liu:2003jj}, we obtain the same 148~fb cross
section from the direct singlet contribution.  We also find 
that the direct octet contribution is about three percent of 
the singlet contribution.  Since the octet NRQCD matrix element
to J/$\psi$, $\langle{}^3S_8\rangle = 
 \langle {\cal O}^{{\rm J/}\psi}_8({}^3_1)\rangle = 
  1.06\times 10^{-2}$~GeV$^3$, is about 100 times smaller than 
corresponding singlet matrix element, it is clear that before
multiplying by the NRQCD matrix elements, as shown in
Fig.~\ref{loxsec}, the production rate for a color
octet active heavy quark pair is larger than the rate of 
producing a singlet pair everywhere in the phase space including 
the region when the invariant mass $s_3$ or $s_4$ is small.  
The octet production rate is much higher than the singlet
rate when both $s_3$ and $s_4$ are large because the active 
quark and the antiquark of the singlet contribution at this order 
cannot come from a pair of heavy quarks that originate from 
either the same virtual photon or gluon. 
With an additional enhancement factor from the color transfer
mechanism, the octet contribution will also peak in the
region where the invariant, $s_3$ or $s_4$, is small.
At the same time, the enhancement in the color octet
coefficient functions away from these regions seems to
imply that the cross section for color transfer will decay
less rapidly than might be expected on the basis of
the explicit factors of $\bar\beta_S$ alone in
Eq.\ (\ref{colortransfersigma}).    

\begin{figure}[t!]
\begin{center} 
\psfig{file=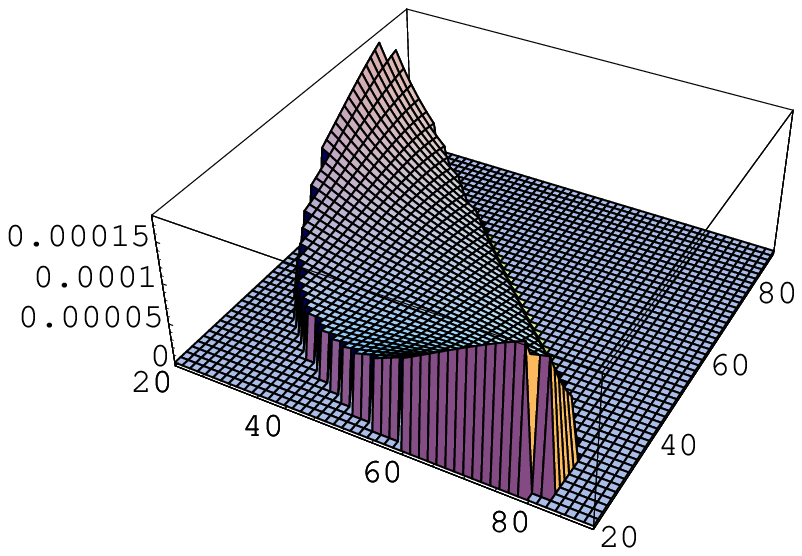,width=2.5in,angle=0}
\hskip 0.3in
\psfig{file=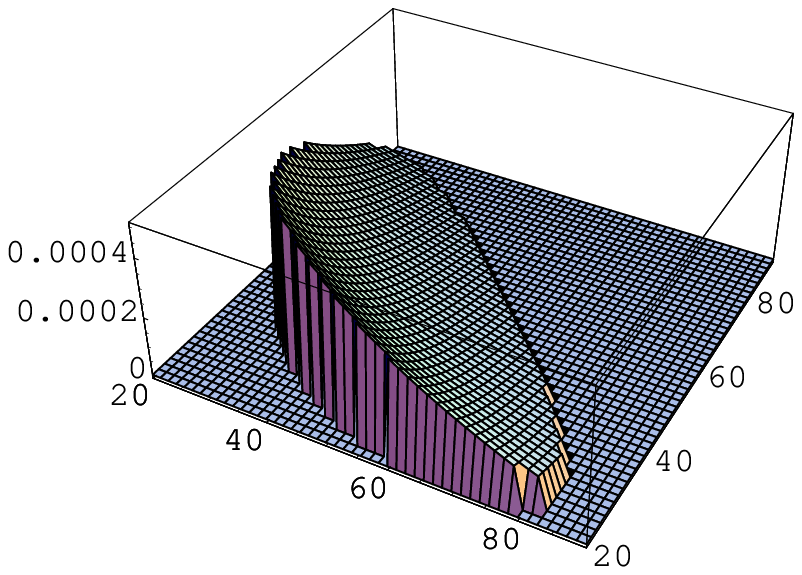,width=2.5in,angle=0}
\caption{The integrand,
in pb/GeV$^3$, of the $s_3\,s_4$ integration  for the production
rate of an active heavy quark pair, defined in 
Eq.~(\protect\ref{lowestsigmahat}), with the pair in a
color singlet (left) or a color octet (right) state.
}
\label{loxsec}
\end{center} 
\end{figure}

\section{High Energy Behavior and Fragmentation}

Equation (\ref{colortransfersigma}) estimates the effect of color
transfer on the production cross section as an energy-independent 
factor times a perturbative cross section.
In this section, we show why this is a natural assumption, even though
the color transfer process acts in a restricted
region of  phase space, specified by Eq.\ (\ref{vordered}).  
It is necessary to check that 
color transfer survives in the high energy limit, and that it can
play a role in the fragmentation of heavy flavor 
at high transverse momentum.

\subsection{Color transfer in fragmentation}

\begin{figure}
\begin{center}
\epsfig{figure=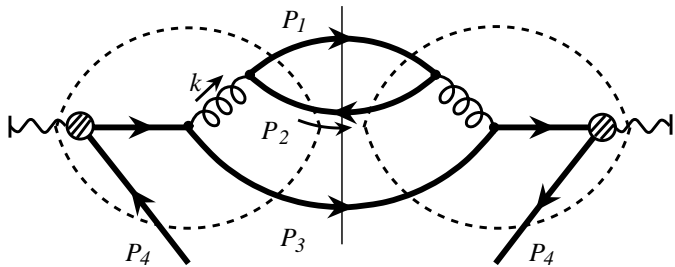,width=3.5in}
\caption{Representative fragmentation contribution
to associated production, discussed in the text.   
The dashed lines represent contributions that would
be included in the short-distance function at low energies, 
as in Fig.\ \ref{nlofig}.}
\label{hienfig}
\end{center}
\end{figure}

Figure \ref{hienfig} shows a typical fragmentation process
for associated production at lowest order, in cut diagram
notation.  To be definite, we discuss 
fragmentation in leptonic annihilation, as in Fig.\ \ref{lofig}.
The underlying factorization that justifies  fragmentation analysis extends as
well to hadronic scattering, as shown in Ref.\ \cite{Nayak:2006fm}.

The shaded circles in Fig.\ \ref{hienfig} represent 
as above short-distance functions,
now with all propagators off-shell by  $Q^2$, 
where $Q$ is the center of mass energy.   
We assume all $P_i^2=m^2$, taking both heavy flavor masses equal.
All propagators enclosed within the dashed lines are then off-shell by
at least $4m^2$, and remain short-distance contributions
from the point of view of NRQCD, as in Fig.\ \ref{nlofig}.

From our previous considerations, we know that
the color transfer process arises from truly
infrared gluons, and will thus not affect the propagators
of lines that are off-shell by the scales $m$ or $Q$.
Rather,  the infrared enhancement will
be associated with factors that  depend sensitively
on the relative velocities of the external lines.  
We will now show that color transfer is leading
power in $Q$, although it will not appear in logarithmic
enhancements associated with large relative
velocities between $P_3$ and $P_1$ or $P_2$.

Let us denote by $d\tau_n(Q^\mu)$ differential $n$-particle phase space
at total four-momentum $Q^\mu$.   
The inclusive associated-pair production cross section
at total momentum $Q^\mu\equiv (Q,\vec{0})$ can then be represented as
\bea
\int d\tau_4(Q^\mu) \frac{d\sigma(\{P_i\})}{d\tau_4}
&=& \frac{1}{2Q^2}\,
\int \frac{d^3\vec{P}_4}{2(2\pi)^3\sqrt{\vec{P}_4^{\; 2}+m^2}}\ 
d\tau_3\left(Q^\mu-P^\mu_4\right)\
\left | \, M_4(Q,P_1\dots P_3)\, \right |^2
\nonumber\\
&\ & \hspace{-20mm} = \frac{1}{8(2\pi)^3Q^3}\, \int d\Omega_4\, \int ds_{123}\, 
\sqrt{ \left( \frac{Q^2+m^2-s_{123}}{2Q} \right)^2 - m^2}\
\int d\tau_3\left(Q^\mu-P^\mu_4 \right)\
\left | \, M_4(Q,P_1\dots P_3)\, \right |^2\, .
\nonumber\\
\label{4to3}
\eea
Here and below, we introduce the notation 
$s_{ij\dots} \equiv (P_i+P_j+\dots)^2$,
while $M_n$ is the  amplitude for the production of $n$ heavy
particles. A sum over final-state
spins is assumed.  The vector $P_4$ is on-shell, 
with the direction of its spatial
momentum specified by the angular integral, $\int d\Omega_4$.

At high energy and fixed invariant mass $\sqrt{s_{123}}$ of the
three-particle (1,2,3) system, the amplitude factorizes into a hard
function for single-pair production multiplied by an integral
that describes the fragmentation of a parent (anti)quark into
an (anti)quark plus a pair, as in Fig.\ \ref{hienfig}.     In the high-energy
limit for fixed $s_{123}$, the corresponding squared amplitude integrated over
three-particle phase space then factorizes into the squared amplitude for
two-particle production, times logarithmic integrals characteristic 
of fragmentation.   Suppressing dependence on the relative
angles of the $\vec{P}_i$, $i=1,2,3$, the remaining integrals in three-particle
phase space appear in the factorized expression as
\bea
\int d\tau_3\left(Q^\mu-P^\mu_4\right)\ 
\left | \, M_4(Q,P_1\dots P_3)\, \right |^2 
&\sim& |\, M_2(P_4,\overline{Q-P_4})\, |^2\
 \int d\tau_3(Q^\mu-P_4^\mu)\, \frac{1}{s_{123}s_{12}}
\nonumber\\
&\ & \hspace{-40mm} \sim\  M_2(P_4,\overline{Q-P_4})\, |^2\
\frac{1}{s_{123}^2}
\ \int_{4m^2} ds_{23}\ \frac{ds_{12}}{s_{12}}\; 
\theta(s_{123}-s_{12}-s_{23}-3m^2)\, ,
\label{3pps}
\eea
where $\overline{Q-P_4}$ denotes an on-shell four-vector 
recoiling against $P_4$.  
The kinematic dependence in the first line of
(\ref{3pps}) comes from the denominators of the squared
propagators shown in Fig.\ \ref{hienfig},
which give $1/s_{12}^2s_{123}^2$, multipied by 
momentum factors from the spinor algebra, 
which give an additional factor in the numerator of $s_{12}s_{123}$.
A remaining, overall factor of $\gamma\cdot(Q-P_4)$ has been
absorbed into the squared two-particle matrix element
$|M_2|^2$ in the expression.
In these terms, three-particle phase space reduces to an infrared finite, but
logarithmic, integral over the squared invariant mass of the (1,2) pair,
times a free integral over the squared invariant mass of the
(2,3) pair.  
The theta function restriction on  $s_{12}$ and $s_{23}$
reflects an identity for three-particle phase space:
$s_{123}=s_{12}+s_{23}+s_{13}+3m^2$.

At lowest order (Fig.\ \ref{hienfig}, for example), 
the $s_{12}$ integral in Eq.\ (\ref{3pps})
 behaves as  $s_{123}\ln(s_{123}/4m^2)$.  This leads in turn to
an additional logarithmic integral over $s_{123}$ in Eq.\ (\ref{4to3}).
These are standard logarithmic enhancements
associated with the evolution of a fragmentation function.
If we demand, however, that one of the two pairs  of Fig.\ \ref{hienfig} form
a heavy quarkonium, we will restrict either $s_{23}$, for the 
lowest-order color singlet contribution,
or $s_{12}$, for the lowest-order color octet, to remain close to $4m^2$.
The available phase space for the formation of the heavy quarkonium
is set by the requirement that the active pair should have
relative momentum of the order $mv$.
This limitation on phase space is taken into account 
in NRQCD matrix elements, which always include a
scaling of order $v^3$.   
In lowest-order singlet production, this limitation would be placed on
$s_{23}$, which eliminates both of the logarithmic 
enhancements identified above.
For the lowest-order octet mechanism,   
the restriction can be placed on $s_{12}$, and
we may still generate a single logarithm from the integral over $s_{123}$.

We can now consider infrared-sensitive NNLO 
color transfer corrections to Fig.\ \ref{hienfig}.
In this case, we generate additional dependence
 on the relative velocities of the
heavy particles, and hence on the invariants $s_{23}$ and $s_{12}$.
As we have seen, this dependence arises from the square of 
the NLO correction given in Eq.\ (\ref{firstorder}),
and it decreases rapidly with
increasing relative velocities between the
active pair and the spectator.
Thus, such a correction suppresses the cross section when $s_{123}$ grows to
a multiple of $9m^2$.   

To be specific, recalling the definition of the variable $\bar\beta_S$
 in Eqs.\ (\ref{barbetadef}) and (\ref{betaSdef}), and neglecting $v$
 compared to $\bar\beta_S$, we find an additional $s_{123}$-dependence
 in the integrand of Eq.\ (\ref{3pps}),
\bea
 \frac{(1- \bar\beta_S^2)^{3}}{\bar\beta_S^4}
 \int ds_{23}\ \frac{ds_{12}}{s_{12}}\;
 \theta(s_{123}-s_{12}-s_{23}-3m^2) 
&\ & 
\nonumber\\
&\ & \hspace{-60mm} =\
\left[\,  \frac{64m^6}
               {(s_{123}-9m^2)(s_{123}-5m^2)(s_{123}-m^2)}\,
\right]^2\, 
 \int ds_{23}\ \frac{ds_{12}}{s_{12}}\; 
\theta(s_{123}-s_{12}-s_{23}-3m^2)\, ,
\eea
where the lower limit of the $s_{123}$ integral
will be set by $\bar\beta_S > v$, and where
there is a rapid decrease for $s_{123}\gg m^2$.
As with the color octet mechanism for this lowest-order process,
the (1,2) pair forms the bound state, and
we anticipate an overall suppression in the $s_{12}$ integral
of $v^3$ on dimensional grounds.
Although non-logarithmic, the $s_{123}$ integral remains 
leading power in $Q$,
with no further suppression by powers of $v$.

\subsection{Light quarks and polarization}

By referring to Fig.\ \ref{hienfig}, we can make a very simple
observation relevant to the comparison of the 
associated production cross section for
heavy quarkonia with heavy flavors compared to
light flavors only.   The leading power
behavior for the production of single heavy quark pairs in close
proximity in phase space is due entirely to fragmentation.
The color transfer contribution for fragmentation diagrams as
shown in the figure is present only for associated production,
since light quarks will unavoidably have much larger
relative velocities, $\bar\beta_S$.   At  lowest order, then,
we would expect the fragmentation of a light quark to a heavy
quarkonium to proceed only through the color octet
matrix element.    For leptonic annihilation in particular,
where gluon fragmentation is yet higher order in $\alpha_s$, 
the color transfer
process affects heavy-flavor associated production only.   

Related considerations apply to the prediction of transverse
polarizations for vector heavy quarkonia produced from
gluon fragmentation \cite{cdfpolar,thypolar,Baek:1997ms}.   
It is easy to check that the
color transfer process does not give such a prediction.
The color transfer itself respects the spin of the pair,
which reflects the polarization of a collinear gluon
in the fragmentation process (the gluon of invariant
mass $s_{12}$ in Fig.\ \ref{hienfig}).  
Indeed, at high energies, the polarization of this
gluon naturally includes large longitudinal components.
A further study of this issue is clearly in order.

\section{Summary and Conclusion}

The color transfer process that we have described above
provides a new viewpoint on the hadronization of heavy
quarks in associated production.   In color transfer, 
the pair produced from a single gluon can
be transformed from octet to singlet representation
by the field of an open (anti)quark that is at
sufficiently low relative velocity.  
We can picture it as a process
that catalyzes quarkonium production when another
heavy quark is nearby in phase space.   
Experimental confirmation of such
a process could provide a significant tool to study
dynamical processes in quantum chromodynamics.
Color transfer complements the 
standard picture of hadronization through string breaking,
in which color neutral pairs are formed from quarks and antiquarks
that originate from separate virtual gluons \cite{preconfine}.   
The latter is leading power in the number of colors, 
and color transfer is hence 
nonleading in this expansion.

We have shown that in perturbation theory, color transfer appears first
at NNLO.  At this level, it is positive, but not infrared
safe.  Although nonperturbative it cannot be described 
in terms of the matrix elements
of NRQCD in general. Without experimental input, we are not yet in
the position to give a realistic estimate of its importance.

Color transfer could be part of the explanation of well-known
anomalies in comparisons of data to NRQCD predictions 
\cite{cdfpolar,associated},
and it may be possible to confirm its role through a number of
qualitative predictions that flow naturally from the 
perturbative reasoning above.  

First, we would expect color transfer to
produce phase space distributions
that are  peaked at small relative
invariant mass between the heavy quarkonium and closest open heavy flavor
in associated production.
For example, an experimental signal of this effect would be 
an enhancement of $J/\psi$ 
production relative to NRQCD estimates at low values of
the invariant masses of $J/\psi - D$ and $J/\psi - \bar D$
pairs.   At high energies, a similar effect should
be found both in $\Upsilon-D$ and $\Upsilon-B$ systems.
Second, if color transfer enhances associated production
at large transverse momenta, it could also help
explain the observed polarizations of heavy quarkonia.

A Dalitz plot analysis of quarkonium/open flavor 
final states at B factory energies could show enhancements on the
low pair mass corners.  At the same time, our analysis suggests that in
the interior of such a Dalitz plot, an NRQCD analysis
based on the color singlet mechanism should succeed.
In general, semi-inclusive measurements of the distribution of
heavy flavor in final states could even enable us to probe
the time evolution of flavor, momentum and spin
in the formation of the final state.  This would complement
the inclusive viewpoint which is built into the
important NRQCD calculations of Refs.\ 
\cite{Artoisenet:2007xi,Zhang:2006ay}.

Color transfer is distinct from, yet in
some ways analogous to, the usual NRQCD color octet mechanism,
which relies on soft gluon radiation.
It is kinematically dependent on
soft gluon exchange with other particles in the
final state, and hence is not ``universal" in
the sense of NRQCD.   On the other hand,
the effect  decays with relative transverse momentum,
and hence does not spoil high-$p_T$ factorization,
as described, for example, in \cite{Nayak:2005rt}.
Similarly, it does not occur at all through exchanges
with  massless particles, and hence is specific to 
associated production with heavy flavor.

In conclusion, although many issues remain to be studied, 
the color transfer mechanism suggests a number of phenomenological
signals, which should make it possible to test its relevance
to associated production.   If it does pass these experimental
tests, it may offer  insight into the dynamics of
color  in hadronization.

\subsection*{Acknowledgments}
We thank Geoff Bodwin for many useful discussions on factorization.
This work was supported in part by the National Science Foundation, 
grants PHY-0354776, PHY-0354822 and PHY-0653342, by the U.S.
Department of Energy under Grant Nos.~DE-FG02-87ER40371
and DE-FG02-01ER41195 and Contract DE-AC02-06CH11357, and in part 
by the Argonne University of Chicago Joint Theory Institute (JTI)
Grant 03921-07-137.

\begin{appendix}

\section{Lowest order matrix elements for heavy quarkonium 
associate production}

In this appendix, we provide the leading order squared matrix elements,
$-g_{\mu\nu}W^{\mu\nu}_{(n)}$ 
that appear in Eq.~(\ref{lowestsigmahat}), 
for producing an active pair of heavy quark and 
antiquark of mass $m$, in association with another heavy quark 
pair of the same mass.  For the active pair in a singlet color state,
$n=1$, we have
\begin{eqnarray}
-g_{\mu\nu}W^{\mu\nu}_{(1)}
&=&
\frac{64}{27 m}
\bigg\{
 \frac{2}{
 \left(m^2-s_3\right)^2 
 \left(m^2-s_4\right)^2 
 \left(-2 m^2+s_3+s_4-2 s\right)^2}
\Big[132 m^{10}
     +2 (136 s-101 (s_3+s_4)) m^8 
\nonumber \\
&& \hskip 0.2in   
   +4
   \left(7 s_3^2+108 s_4 s_3+7 s_4^2+240 s^2-58 (s_3+s_4) s\right)
    m^6
\nonumber \\
&& \hskip 0.2in   
   -
   \left(31 s_3^3+149 s_4 s_3^2+149 s_4^2 s_3+31
         s_4^3-380 s^3+316 (s_3+s_4) s^2
    -9 \left(9 s_3^2-2 s_4 s_3+9 s_4^2\right) s 
   \right) m^4
\nonumber \\
&& \hskip 0.2in   
   +
   \left(s_3^4+56 s_4 s_3^3+98 s_4^2 s_3^2+56 s_4^3
         s_3+s_4^4+32 s^4-60 (s_3+s_4) s^3
   \right.
\nonumber \\
&& \hskip 0.5in  
   \left. 
    +12 
    \left(3 s_3^2-2 s_4 s_3+3 s_4^2\right)
     s^2-(s_3+s_4)
    \left(s_3^2-10 s_4 s_3+s_4^2\right) s
   \right) m^2
\nonumber \\
&& \hskip 0.2in   
   +s_3 s_4 
   \left(-9 s_3^3-25 s_4 s_3^2-25 s_4^2 s_3-9
     s_4^3-4 s^3+4 (s_3+s_4) s^2
   + \left(9 s_3^2+14 s_4 s_3+9 s_4^2 \right) s
   \right)
\Big]
\nonumber \\
&&
- \frac{1}{\left(m^2-s_3\right)^4}
\Big[
  64 m^6+(41 s_3+9 s_4+47 s) m^4+2 s_3
      (s-5 (s_3+s_4)) m^2+s_3^2 (s_3+s_4-s)
\nonumber \\
&&
\hskip 0.2in
  +
  \frac{2\left(m^2-s_3\right)}{-2 m^2+s_3+s_4-2 s}
  \big[ 257 m^6+3 (2 s_3-7 s_4+54 s) m^4
\nonumber \\
&&
\hskip 0.5in
    -\left(7 s_3^2+36 s_4 s_3+10 s_4^2-16 s^2
          + 2 (s_3+7 s_4) s\right) m^2
   +s_3 s_4 (s_3+2 s_4-2 s)
  \big]
\nonumber \\
&&
\hskip 0.2in
  +
  \frac{4 \left(m^2-s_3\right)^2}
       {\left(-2 m^2+s_3+s_4-2 s\right)^2}
  \big[ 160 m^6+3 (5 s_3-15 s_4+37 s) m^4
\nonumber \\
&&
\hskip 0.5in
    -2 \left(5 s_3^2+14 s_4 s_3-2 s s_3
            +7 s_4^2-4 s^2+5 s_4 s\right) m^2
   +\left(2 s_3^2+4 s_4 s_3+3 s_4^2\right) (s_3+s_4-s)
  \big]
\Big]
\\
&&
-\frac{1}{\left(m^2-s_4\right)^4} 
\Big[
64 m^6+(9 s_3+41 s_4+47 s) m^4+2 s_4 
      (s-5 (s_3+s_4)) m^2+s_4^2
       (s_3+s_4-s)
\nonumber \\
&&
\hskip 0.2in
  +
  \frac{2\left(m^2-s_4\right)}{-2 m^2+s_3+s_4-2s}
  \big[ 257 m^6+(6 (s_4+27 s)-21 s_3)m^4
\nonumber \\
&&
\hskip 0.5in
   -\left(10 s_3^2+36 s_4 s_3+7 s_4^2-16 s^2+2 (7 s_3+s_4) s\right)
     m^2+s_3 s_4 (2 s_3+s_4-2 s)
  \big]
\nonumber \\
&&
\hskip 0.2in
  +
  \frac{4\left(m^2-s_4\right)^2}
       {\left(-2 m^2+s_3+s_4-2 s\right)^2}
  \big[ 160 m^6+(-45 s_3+15 s_4+111 s) m^4
\nonumber \\
&&
\hskip 0.5in
        -2\left(7 s_3^2+14 s_4 s_3+5s s_3+5 s_4^2-4 s^2-2
           s_4 s\right) m^2
        +\left(3 s_3^2+4 s_4 s_3+2s_4^2\right)
        (s_3+s_4-s)
  \big]
\Big]
\bigg\} \, .
\nonumber
\end{eqnarray}
For the pair in an octet color state,
$n=8$, we have
\begin{eqnarray}
-g_{\mu\nu}W^{\mu\nu}_{(8)}
&=&
\frac{1}{9 m
   \left(m^2-s_3\right)^4
   \left(m^2-s_4\right)^4}
\bigg\{
   1594 m^{14}-4588 s_3 m^{12}-4588
   s_4 m^{12}+4285 s_3^2 m^{10}+4285
   s_4^2 m^{10}
\nonumber \\
&& \hskip 0.2in
   +13670 s_3 s_4
   m^{10}-1961 s_3^3 m^8-1961 s_4^3
   m^8-12723 s_3 s_4^2 m^8-12723
   s_3^2 s_4 m^8+553 s_3^4 m^6+553
   s_4^4 m^6
\nonumber \\
&& \hskip 0.2in
   +5699 s_3 s_4^3
   m^6+10316 s_3^2 s_4^2 m^6+5699
   s_3^3 s_4 m^6-75 s_3^5 m^4-75
   s_4^5 m^4-1417 s_3 s_4^4
   m^4-3760 s_3^2 s_4^3 m^4
\nonumber \\
&& \hskip 0.2in
   -3760
   s_3^3 s_4^2 m^4-1417 s_3^4
   s_4 m^4+127 s_3 s_4^5 m^2+791
   s_3^2 s_4^4 m^2+846 s_3^3
   s_4^3 m^2+791 s_3^4 s_4^2
   m^2+127 s_3^5 s_4 m^2
\nonumber \\
&& \hskip 0.2in
   -61 s_3^2
   s_4^5-137 s_3^3 s_4^4-137
   s_3^4 s_4^3-61 s_3^5
   s_4^2
\nonumber \\
&& 
+\frac{4 \left(m^2-s_3\right)\left(m^2-s_4\right)}
      {-2 m^2+s_3+s_4-2 s} 
\Big[
   478 m^{12}-1130
   (s_3+s_4) m^{10}+\left(619
   s_3^2+3084 s_4 s_3+619
   s_4^2\right) m^8
\nonumber \\
&& \hskip 0.2in
   -4 (s_3+s_4)
   \left(74 s_3^2+383 s_4 s_3+74
   s_4^2\right) m^6+2 \left(84 s_3^4+256
   s_4 s_3^3+441 s_4^2
   s_3^2+256 s_4^3 s_3+84
   s_4^4\right) m^4
\nonumber \\
&& \hskip 0.2in
   -(s_3+s_4)
   \left(19 s_3^4+110 s_4 s_3^3+40
   s_4^2 s_3^2+110 s_4^3
   s_3+19 s_4^4\right) m^2
\nonumber \\
&& \hskip 0.2in
   +s_3
   s_4 \left(7 s_3^4+21 s_4
   s_3^3+6 s_4^2 s_3^2+21
   s_4^3 s_3+7
   s_4^4\right)
\Big]
\\
&& 
+\frac{4 \left(m^2-s_3\right)^2 \left(m^2-s_4\right)^2}
      {\left(-2 m^2+s_3+s_4-2 s\right)^2}
\Big[
   244 m^{10}-458
   (s_3+s_4) m^8+4 \left(17
   s_3^2+304 s_4 s_3+17
   s_4^2\right) m^6
\nonumber \\
&& \hskip 0.2in
   -2 (s_3+s_4)
   \left(57 s_3^2+136 s_4 s_3+57
   s_4^2\right) m^4+2 \left(19 s_3^4+52
   s_4 s_3^3+52 s_4^2
   s_3^2+52 s_4^3 s_3+19
   s_4^4\right) m^2
\nonumber \\
&& \hskip 0.2in
   -(s_3+s_4)
   \left(3 s_3^4+8 s_4 s_3^3+12
   s_4^2 s_3^2+8 s_4^3
   s_3+3 s_4^4\right)
\Big]
\nonumber \\
&& 
+\frac{2 \left(7 m^2-3 s_3\right) \left(m^2-s_3\right)^2
         \left(7 m^2-3 s_4\right) \left(m^2-s_4\right)^2 s^2}
      {m^2}
+\frac{9 s_3^3 s_4^3 \left(6 m^4+2 (s_3+s_4) m^2+s_3^2+s_4^2\right)}
      {m^2}
\nonumber \\
&& 
+\frac{2 s}{m^2}
\Big[
   252 m^{14}-677 (s_3+s_4)
   m^{12}+\left(301 s_3^2+2438 s_4
   s_3+301 s_4^2\right)
   m^{10}
\nonumber \\
&& \hskip 0.2in
   +(s_3+s_4) \left(173
   s_3^2-2185 s_4 s_3+173
   s_4^2\right) m^8+\left(-97 s_3^4+410
   s_4 s_3^3+1946 s_4^2
   s_3^2+410 s_4^3 s_3-97
   s_4^4\right) m^6
\nonumber \\
&& \hskip 0.2in
   +s_3 s_4
   (s_3+s_4) \left(33 s_3^2-581
   s_4 s_3+33 s_4^2\right)
   m^4+s_3^2 s_4^2 \left(25
   s_3^2+166 s_4 s_3+25
   s_4^2\right) m^2-9 s_3^3 s_4^3
   (s_3+s_4)
\Big]
\bigg\}\, .
\nonumber
\end{eqnarray}
 The invariants in these expressions are
 defined as $s=(P+P_3+P_4)^2$, $s_3=(P+P_3)^2$, 
and $s_4=(P+P_4)^2$, respectively.  

The phase space integration of $ds_3\, ds_4$ 
in Eq.~(\ref{lowestsigmahat}) is given by
the condition, $\Phi(s,s_3,s_4,m)\ge 0$, with the
function \cite{Davydychev:2003cw}
\begin{eqnarray}
\Phi(s,s_3,s_4,m)
= s_3 s_4 (6 m^2 - s_3 - s_4 + s)
+ 3 (s-m^2)(s_3+s_4) m^2 - (s+6m^2)(4s+m^2) m^2 
+ 2( 9 s + 4 m^2) m^4 \, .
\end{eqnarray}

\end{appendix}


\end{document}